\documentclass[11pt,a4paper,english]{article}
\usepackage{palatino}
\usepackage[T1]{fontenc}
\usepackage[latin1]{inputenc}
\usepackage{graphicx}
\usepackage{amssymb}

\makeatletter

\newcommand{\boldsymbol}[1]{\mbox{\boldmath $#1$}}

 \usepackage{amsthm, amsmath, amsfonts, amssymb}
 \theoremstyle{plain}    
 \newtheorem{thm}{Theorem}
 \theoremstyle{definition}
 \newtheorem{defn}[thm]{Definition}
 \theoremstyle{plain}    
 \newtheorem{lem}[thm]{Lemma} 
 \theoremstyle{plain}    
 \newtheorem{prop}[thm]{Proposition} 

\usepackage{floatflt}

\AtBeginDocument{
  
}

\usepackage{babel}
\makeatother
\begin{document}

\title{A semi-classical K.A.M. theorem}

\author{Nicolas Roy%
\footnote{Geometric Analysis Group, Institut für Mathematik, Humboldt Universität,
Rudower Chaussee 25, Berlin D-12489, Germany. Email: \texttt{roy@math.hu-berlin.de}%
}~%
\footnote{The author would like to thank Hermann Schulz-Baldes for his numerous
comments about the manuscript.%
}~%
\footnote{This paper has been partially supported by the European Commission
through the Research Training Network HPRN-CT-1999-00118  \char`\"{}Geometric
Analyis\char`\"{}.%
}}

\maketitle
\begin{abstract}
We consider a semi-classical completely integrable system defined
by a $\hbar$-pseudodifferential operator $\hat{H}$ on the torus
$\mathbb{T}^{d}$. In order to study perturbed operators of the form
$\hat{H}+\hbar^{\kappa}\hat{K}$, where $\hat{K}$ is an arbitrary
pseudodifferential operator and $\kappa>0$, we prove the conjugacy
to a suitable normal form. This is then used to construct a large
number of quasimodes. 
\end{abstract}

\section{Introduction}

\subsection{Semi-classical perturbations of completely integrable systems}

Let us start with a short overview of the context in which this paper
takes place. First of all, one knows from the Mineur-Arnol'd-Liouville
Theorem \cite{mineur,arnold_1} that on a symplectic manifold $\mathcal{M}$,
a Hamiltonian $H$ admitting a momentum map is completely integrable
(CI in short) in the sense that $\mathcal{M}$ is fibered almost everywhere
by invariant lagrangian tori on which the dynamics of $H$ is very
simple, namely conjugate to translations on the standard affine torus
$\mathbb{T}^{d}$. On each torus, the trajectories are thus either
periodic or quasi-periodic.

Now, one knows from Poincaré \cite{poincare_123} that a generic perturbation
$H+\varepsilon K$ will destroy its complete integrability. Nevertheless,
the celebrated K.A.M. Theorem \cite{kolmogorov,arnold_2,moser_1}
insures that a {}``large part of the CI character'' survives after
a small perturbation $\varepsilon K$ is added. Namely, the tori on
which the completely integrable dynamics satisfies a certain diophantine
relation are simply slightly deformed into invariant lagrangian tori
of $H+\varepsilon K$ without being destroyed by the perturbation.
They are called \emph{K.A.M. tori}. 

On the other hand, in the context of semi-classical analysis, several
authors \cite{colin_5,charbonnel} transposed to the pseudodifferential
operators (PDOs in short) the Mineur-Arnol'd-Liouville Theorem. Thanks
to these works, any PDO which admits a semi-classical momentum map
can be conjugate to an operator on $T^{*}\mathbb{T}^{d}=\left\{ \left(x,\xi\right)\mid x\in\mathbb{T}^{d},\xi\in\mathbb{R}^{d}\right\} $
with a symbol depending only on $\xi$, microlocally in a neighborhood
of any connected component of any compact regular fiber of the momentum
map. We will thus work from the beginning in the angle-action coordinates
and use a well-adapted pseudodifferential calculus. Actually, because
of the very particular structure of the torus $\mathbb{T}^{d}$, one
is able to construct a pseudodifferential calculus involving globally
defined (total) symbols. These operators are sometimes called {}``\emph{periodic
PDO}s'' in the literature. Such PDOs have been studied by several
authors \cite{coifman_meyer,dunau,turunen_vainikko} but always without
a small parameter. In this paper, we use a $\hbar$-version of these
theories.

We thus begin with a PDO $\hat{H}$ with symbol $H\left(\xi\right)$
being a CI Hamiltonian in the classical sense. It is easy to see that
its spectrum is $\left\{ H\left(\hbar k\right)\mid k\in\mathbb{Z}^{d}\right\} $
and the associated eigenvectors are simply $e^{ikx}$. Now, any perturbation
of $\hat{H}$ naturally relies on two parameters : a parameter $\varepsilon$
which controls the {}``intensity'' of the perturbation and the semi-classical
parameter $\hbar$. In this paper, we will be interested in the regime
$\varepsilon\sim\hbar^{\kappa}$ and thus consider perturbed operators
of the form $\hat{H}+\hbar^{\kappa}\hat{K}$, with $\kappa>0$ and
$\hat{K}$ any PDO.

As in the {}``classical'' K.A.M. Theorem, one needs to impose a
nondegeneracy condition on the unperturbed Hamiltonian $H\left(\xi\right)$.
For example, one may require the Hessian matrix $\partial_{\xi_{i}}\partial_{\xi_{j}}H$
to be non-degenerate, i.e. Kolmogorov's condition. We will rather
use a weaker one, which will be stated precisely later on. 

The perturbed operator $\hat{H}+\hbar^{\kappa}\hat{K}$ depends on
$\hbar$ and we want to investigate the associated family of spectra
$\sigma_{\hbar}$ depending on $\hbar$. Actually, the use of the
pseudodifferential calculus allows to investigate $\hbar^{\infty}$-quasimodes
rather than genuine eigenvectors. We recall that a quasimode is a
family of functions $\varphi_{\hbar}\in L^{2}\left(\mathbb{T}^{d}\right)$
together with a family of numbers $E_{\hbar}$ depending on $\hbar$,
such that $\left\Vert \varphi_{\hbar}\right\Vert _{L^{2}}=1$ and
\[
\left\Vert \left(\hat{H}+\hbar^{\kappa}\hat{K}-E_{\hbar}\right)\varphi_{\hbar}\right\Vert _{L^{2}}=O\left(\hbar^{\infty}\right).\]
When the operator under consideration is self-adjoint, then $E_{\hbar}$
is $\hbar^{\infty}$-close to the spectrum $\sigma_{\hbar}$%
\footnote{Nevertheless, this does not imply that $\varphi_{\hbar}$ is close
to any eigenvector, as first remarked by Arnol'd \cite{arnold_5}. %
}. The main result of this paper is the construction of a large number
of quasi-modes of $\hat{H}+\hbar^{\kappa}\hat{K}$, as stated below%
\footnote{The precise statements are given in Theorem \ref{theo_quasimode_non_resonant}
and Proposition \ref{prop_volume_bloc}.%
}. 

\begin{thm}
\label{theo_main}Let $\hat{H}$ be a PDO with non-degenerate symbol
$H\left(\xi\right)$ and $\hbar^{\kappa}\hat{K}$ any PDO. Denote
by $\left\langle \left\langle K\right\rangle \right\rangle \left(\xi\right)$
the average over the torus of the symbol $K\left(x,\xi\right)$. For
any fixed $\delta\in\left(0,\frac{\kappa}{3}\right)$, there exists
a {}``quasi-resonant'' zone $\mathcal{Z}\subset\mathbb{R}_{\xi}^{d}$
depending on $\hbar$ and of relative volume\[
vol\left(\mathcal{Z}\right)\sim\hbar^{\delta-\varepsilon},\]
where $\varepsilon$ can be taken arbitrarily small, and such that
for all $k_{\hbar}\in\mathbb{Z}^{d}$ with $\hbar k_{\hbar}\in\mathbb{R}^{d}\setminus\mathcal{Z}$,
there exists a $\hbar^{\infty}$-quasimode $\left(\varphi_{\hbar},E_{\hbar}\right)$
of $\hat{H}+\hbar^{\kappa}\hat{K}$ with\\
$\bullet$ $E_{\hbar}=H\left(\hbar k_{\hbar}\right)+\hbar^{\kappa}\left\langle \left\langle K\right\rangle \right\rangle \left(\hbar k_{\hbar}\right)+O\left(\hbar^{\kappa+\alpha}\right)$,
where $\alpha=\min\left(1-\delta,\kappa-3\delta\right)$\\
$\bullet$ $\varphi_{\hbar}\left(x\right)=e^{ik_{\hbar}x}+O\left(\hbar^{\kappa-1-\delta}\right)$.
\end{thm}
\begin{center}\emph{\includegraphics[%
  width=5cm,
  keepaspectratio]{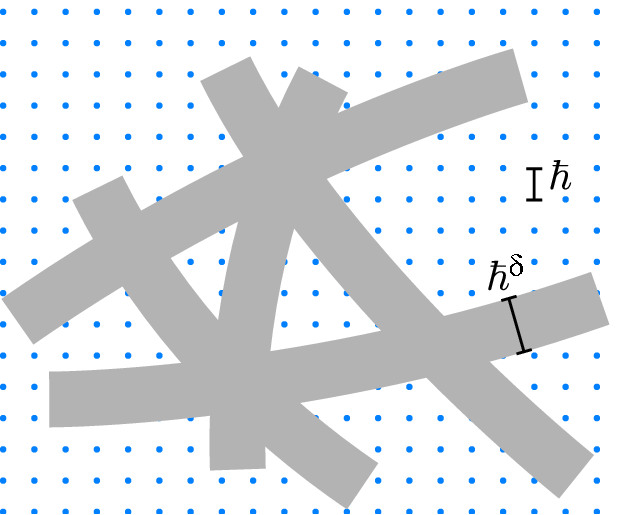}}\end{center}

The picture above represents the case $d=2$ : the quasi-resonant
zone $\mathcal{Z}$ is in grey and the dots represent the lattice
$\hbar\mathbb{Z}^{d}$. The assertion about the {}``relative volume''
means that for any ball $\mathcal{O}\subset\mathbb{R}^{d}$ the volume
of $\mathcal{Z}\cap\mathcal{O}$ is of order $vol\left(\mathcal{Z}\right)\sim\hbar^{\delta-\varepsilon}.$
In the semi-classical limit, the \emph{non-resonant set $\mathbb{R}^{d}\setminus\mathcal{Z}$}
tends to the set of diophantine tori which are preserved by the perturbation,
due to K.A.M. theory. This should be compared to the fact that all
the eigenvalues $H\left(\hbar k_{\hbar}\right)$ of $\hat{H}$ with
$\hbar k\in\mathbb{R}^{d}\setminus\mathcal{Z}$ are only slightly
modified by the perturbation $\hbar^{\kappa}\hat{K}$ in the sense
that there exists a $\hbar^{\infty}$-quasi-eigenvalue $\hbar^{\kappa}$-close
to $H\left(\hbar k_{\hbar}\right)$. The first correction is the average
of the symbol of the perturbation. Hence, this result should be regarded
as a semi-classical K.A.M. theorem.

\subsection{A result by Feldman-Knörrer-Trubowitz.}

Before starting to investigate our problem, we mention the article
\cite{feldman_knorrer_trubowitz_1} by Feldman, Knörrer and Trubowitz
(FKT in short). The authors studied the high energy asymptotics for
the periodic Schrödinger operator $-\Delta+V$ on the torus. Even
though this problem is not in the $\hbar$-pseudodifferential context,
one can use the {}``usual'' correspondence \emph{Semi-classical
limit $\leftrightarrow$ High frequency limit} in order to compare
the results. Starting from the eigenvalue problem $-\Delta\varphi+V\varphi=\lambda\varphi$,
setting $\lambda=\frac{E_{\hbar}}{\hbar^{2}}$ and multiplying everything
by $\hbar^{2}$, the problem becomes \[
-\hbar^{2}\Delta\varphi+\hbar^{2}V\varphi=E_{\hbar}\varphi\]
 and the high energy limit $\lambda\rightarrow+\infty$ corresponds
to investigating eigenvalues $E_{\hbar}$ of order $1$ in the semi-classical
limit $\hbar\rightarrow0$. Under this correspondence, FKT's result
appears as a special case of ours with the symbol of the completely
integrable operator $-\hbar^{2}\Delta$ being $H\left(\xi\right)=\xi^{2}$
and the perturbation being a multiplication operator of order $\hbar^{2}$.

We now recall the relevant result of FKT. Consider the operator $-\Delta+V$
defined on the torus $\mathbb{R}^{d}/\Gamma$ where $\Gamma$ is a
generic lattice of $\mathbb{R}^{d}$, $V$ is a periodic potential
and $d\leq3$. The corresponding unperturbed operator is simply $-\Delta$
and its eigenvectors are $e^{ikx}$ with corresponding eigenvalues
$k^{2}$, for each $k\in\Gamma^{*}$ and where the dual lattice $\Gamma^{*}$
is the Fourier lattice.

\begin{thm}
[FKT]\label{theo_FKT}There exists an {}``exceptional subset''
$S\subset\Gamma^{*}$ of density zero such that for all $k\in\Gamma^{*}\setminus S$
the following holds :\\
$\bullet$ There are 2 eigenvalues $\lambda_{\pm k}$ in the interval\[
\left[k^{2}+\left\langle \left\langle V\right\rangle \right\rangle -\frac{1}{\left|k\right|^{2-\varepsilon}},k^{2}+\left\langle \left\langle V\right\rangle \right\rangle +\frac{1}{\left|k\right|^{2-\varepsilon}}\right],\]
where $\left\langle \left\langle V\right\rangle \right\rangle $ denotes
the average of $V$ over the torus and $\varepsilon>0$ can be taken
arbitrarily small.\\
$\bullet$ The corresponding eigenvectors $\Psi_{\pm k}$ verify \[
\left\Vert \Psi_{\pm k}-\hat{\Pi}\Psi_{\pm k}\right\Vert =O\left(\frac{1}{\left|k\right|^{1-\varepsilon}}\right)\]
where $\hat{\Pi}$ is the projector on the span of $e^{\pm ikx}$.
\end{thm}
The authors call \emph{stable} the unperturbed eigenvalues $k^{2}$
for $k\notin S$ since in the large $k$ limit, they are only slightly
modified when the perturbation is added. The first correction is of
order $O\left(1\right)$ and equals to the average of the {}``perturbation''
$V$, and the next correction is of order $O\left(\frac{1}{\left|k\right|^{2-\varepsilon}}\right)$. 

Our result (Theorem \ref{theo_main}) extends this result to general
CI PDOs, with general pseudodifferential perturbations of any order
$\hbar^{\kappa}$ and with no restriction on the dimension $d$. Moreover,
the \emph{exceptional subset $S$} arising in FKT's result is defined
in quite a tricky way. But it corresponds in our setting to the intersection
of the lattice $\hbar\mathbb{Z}^{d}$ with the quasi-resonant zone
$\mathcal{Z}$ which is quite intuitive and geometric, as we will
see in the sequel.

\subsection{Normal forms and special classes of symbols}

The main tool leading to Theorem \ref{theo_main} is a suitable semi-classical
normal form for the perturbed operator $\hat{H}+\hbar^{\kappa}\hat{K}$,
i.e. a conjugacy of $\hat{H}+\hbar^{\kappa}\hat{K}$ by an unitary
operator $\hat{U}$ : $\hat{U}\left(\hat{H}+\hbar^{\kappa}\hat{K}\right)\hat{U}^{*}=\widehat{NF}+O\left(\hbar^{\infty}\right)$,
where $\widehat{NF}$ will have special properties. The construction
of this normal form is an iterative process whose first step amounts
to looking for a self-adjoint PDO $\hat{P}$ such that the conjugacy
\[
e^{i\hbar^{\kappa-1}\hat{P}}\left(\hat{H}+\hbar^{\kappa}\hat{K}\right)e^{-i\hbar^{\kappa-1}\hat{P}}=\hat{H}+\hbar^{\kappa}\hat{A}+O\left(\hbar^{1+\kappa}\right)\]
 yields a PDO $\hat{A}$ with the {}``simplest'' form as possible,
as we discuss just below. If we write down the corresponding equation
for the symbols, we can check that the cancelation of the lower order
terms is equivalent to solving\[
X_{H}\left(P\right)=K-A,\]
where $X_{H}$ is the Hamiltonian vector field associated with the
CI Hamiltonian $H\left(\xi\right)$ and is thus tangent to each torus
$\mathcal{T}{}_{\xi}=\mathbb{T}^{d}\times\left\{ \xi\right\} $. On
each torus, the solutions of this \emph{homological equation} depend
strongly on the dynamics on the torus (periodic or quasiperiodic).
For example, when the dynamics is periodic, then one can choose $A$
to be the average of $K$ along the trajectories and solve the equation
with a $P$ depending smoothly on $x$. On the other hand, when dynamics
is quasi-periodic and satisfies a diophantine condition, then one
can solve the equation, with $A$ being the average of $K$ over the
whole torus, and obtain a $P$ depending smoothly on $x$.

Unfortunately, one cannot solve this equation in that way, torus by
torus, since CI Hamiltonians are generically \emph{non-degenerate,}
as assumed in Theorem \ref{theo_main}. This property implies in particular
that the vector field $X_{H}$ {}``turns'' when one moves in the
space of tori. In other words, close to each periodic torus lie some
quasi-periodic tori, and vice versa. It is thus impossible to solve
the homological equation torus by torus since this would provide a
function $P\left(x,\xi\right)$ not even continuous with respect to
$\xi$ and thus of course not acceptable as a symbol.

However, it is possible in some sense to {}``interpolate'' between
regions close to periodic motions and regions close to quasi-periodic
motions, and more generally between regions close to any kind of resonance
(non-resonant, partially resonant, periodic...). This can be achieved
by covering the momentum space $\mathcal{B}=\left\{ \xi\right\} $
with \emph{quasi-resonant} regions, i.e. neighborhoods of resonant
tori, as people usually do in Nekhoroshev-like theorems \cite{poeschel_2,benettin_galgani_giorgilli}.
In order for this construction to be of some interest, the involved
neighborhoods must have a thickness which goes to $0$ with $\hbar$,
e.g. of order $\hbar^{\delta}$, with $\delta>0$. But on the other
hand, this forces us to consider symbols $P\left(x,\xi\right)$ whose
dependence on $\xi$ becomes {}``bad'' in the semi-classical limit
$\hbar\rightarrow0$. In fact, we will see that it is possible to
solve the homological equation in the class $\Psi_{\delta}$ of symbols
satisfying \[
\left|\partial_{x}^{\alpha}\partial_{\xi}^{\beta}\left(P_{\hbar}\left(x,\xi\right)\right)\right|\leq C_{\alpha,\beta}\hbar^{m-\delta\left|\beta\right|}.\]
These symbols are actually similar to those used by Sjöstrand in his
study of the semi-excited states \cite{sjostrand}, and one can show
that they indeed form an acceptable class of symbols provided $\delta<1$.

\subsection{Plan of the paper}

In the next section, we give without proof the basic results concerning
the pseudodifferential calculus on the torus with symbols in the class
$\Psi_{\delta}$. Namely, we give the composition law (Moyal's product),
the $L^{2}$-continuity (Calder\'on-Vaillancourt) theorem, the properties
of adjoints and the functional calculus for these PDOs, and we refer
to \cite{roy_1} for detailed proofs.

Section \ref{sec_geom_resonance} is devoted to the construction of
the announced covering of the momentum space $\mathcal{B}=\left\{ \xi\right\} $
by quasi-resonant zones with thickness of order $\hbar^{\delta}$.
This construction is then used in Section \ref{sec_quasi_reson_averaging}
in order to define the notion of \emph{quasi-resonant averaging} of
functions. Roughly speaking, this permits to interpolate between regions
with different kinds of resonance, and thus provides a function which
has a different average property in each of these regions.

Equipped with these tools we are then able to study the perturbed
operator $\hat{H}+\hbar^{\kappa}\hat{K}$ using the semi-classical
normal form of Theorem \ref{theo_quasi_resonant_NF}, in which the
symbol in the normal form is the quasi-resonant average of some symbol
related to $K$. This theorem is based on Proposition \ref{prop_homological_eq}
which insures that one can solve the homological equation, arising
at each step of the normal form iteration, in the class of symbols
$\Psi_{\delta}$ previously defined.

Finally, as an application of this normal form, we show in Theorem
\ref{theo_quasimode_non_resonant} how to build a large number of
quasimodes for the perturbed operator.

\section{\label{sec_PDO_torus}Pseudodifferential operators on the affine
torus}

\subsection{Classes of symbols for periodic $\hbar$-PDOs}

We consider the cotangent bundle $T^{*}\mathcal{T}$ of the $d$-dimensional
affine torus $\mathcal{T}=\left(\mathbb{R}/2\pi\mathbb{Z}\right)^{d}$
and we will denote by $\left(x,\xi\right)$ the canonical variables.
In the following, we denote by $\Lambda^{*}=\mathbb{Z}^{d}$ the lattice
of the Fourier variables, which is the $2\pi$-dual lattice of the
lattice of horizontal $1$-periodic constant vector fields $\Lambda=2\pi\mathbb{Z}^{d}$,
i.e. vector fields of the form $X=X^{j}\frac{\partial}{\partial x^{j}}$
with $X^{j}\in2\pi\mathbb{Z}$ for all $\left(x,\xi\right)$. We will
often denote by $k$ the Fourier variables and by $\tilde{f}\left(k,\xi\right)$
the Fourier series with respect to $x$ of a function $f\left(x,\xi\right)$.

\begin{defn}
\label{def_symbol_sjostrand}Let $m$ and $\delta\geq0$ be two real
constants and $\mathcal{S}\subset T^{*}\mathcal{T}$ any subset. The
\textbf{class of symbols} $\boldsymbol{\Psi_{\delta}^{m}}\left(\mathcal{S}\right)$
is the set of $\hbar$-families of functions $P_{\hbar}\left(x,\xi\right)\in C^{\infty}\left(T^{*}\mathcal{T},\mathbb{C}\right)$,
for $\hbar\in\left(0,1\right]$, such that for all multi-indices $\alpha,\beta\in\mathbb{Z}^{d}$,
there exists a constant $C_{\alpha,\beta}>0$ such that for each point
$\left(x,\xi\right)\in\mathcal{S}$ and each $\hbar\in\left(0,1\right]$,
we have the following upper bound \[
\left|\partial_{x}^{\alpha}\partial_{\xi}^{\beta}\left(P_{\hbar}\left(x,\xi\right)\right)\right|\leq C_{\alpha,\beta}\hbar^{m-\delta\left|\beta\right|}.\]

\end{defn}
When $\mathcal{S}=T^{*}\mathcal{T}$, we simply denote $\Psi_{\delta}^{m}=\Psi_{\delta}^{m}\left(T^{*}\mathcal{T}\right)$.
As well, the class of usual symbols (i.e. for $\delta=0$) is simply
denoted by $\Psi^{m}=\Psi_{0}^{m}\left(T^{*}\mathcal{T}\right)$.
Moreover, it follows from the definition that $\Psi_{\delta}^{m}=\hbar^{m}\Psi_{\delta}^{0}$.
On the other hand, the reader should keep in mind that when $\delta\neq0$
those symbols may not have any well-defined principal symbol $\lim_{\hbar\rightarrow0}P_{\hbar}$.

We denote by $\hat{\Psi}_{\delta}^{m}$ the corresponding class of
PDO's, and by $\hat{P}$ the (left) quantization of a symbol $P$.
A PDO $\hat{P}$ is called a \textbf{negligible} \textbf{operator}
if $\left\Vert \hat{P}\right\Vert _{\mathcal{L}\left(L^{2}\right)}=O\left(\hbar^{\infty}\right)$.
We denote this by$\hat{P}=O\left(\hbar^{\infty}\right).$ We say that
two operators $\hat{A},\hat{B}\in\hat{\Psi}_{\delta}^{m}$ are \textbf{equivalent}
if they satisfy$\hat{A}-\hat{B}=O\left(\hbar^{\infty}\right)$ and
we denote this by $\hat{A}\cong\hat{B}$. Note that this is slightly
weaker than requiring the difference to be in $\bigcup_{m}\Psi_{\delta}^{m}$.

It is also convenient to have a criterion for a function to be in
the class $\Psi_{\delta}^{m}$ expressed in terms of its Fourier series
with respect to the $x$ variable. 

\begin{lem}
\label{lem_OPD_fourier_symbole}A function $P_{\hbar}\left(x,\xi\right)$
is a symbol in the class $\Psi_{\delta}^{m}$ if and only if its Fourier
series $\tilde{P}_{\hbar}\left(k,\xi\right)$ satisfies the following
estimate. For each multi-index $\beta\in\mathbb{Z}^{d}$ and each
positive integer $s$, there exists a constant $C\left(s,\beta\right)>0$
such that for all $k\in\Lambda^{*}$, all $\xi\in\mathcal{B}$ and
all $\hbar\in\left(0,1\right]$, \[
\left|\partial_{\xi}^{\beta}\tilde{P}_{\hbar}\left(k,\xi\right)\right|\leq C\left(s,\beta\right)\frac{\hbar^{m-\delta\left|\beta\right|}}{\left(1+\left|k\right|^{2}\right)^{\frac{s}{2}}}.\]

\end{lem}

\subsection{Composition, $L^{2}$ continuity and functional calculus}

\subsubsection{Asymptotic expansions}

First of all, due to the presence of $\hbar^{-\delta}$ factors arising
in the derivatives of symbols belonging to the class $\hat{\Psi}_{\delta}^{m}$,
one is forced to consider more general asymptotic expansions than
the usual ones (which read $\hbar^{0}P_{0}\left(x,\xi\right)+\hbar^{1}P_{1}\left(x,\xi\right)+...$
and which are sometimes called \emph{classical symbols}). 

\begin{defn}
\label{def_dev_asymptotic}Let $\delta\geq0$, $m$ and $\alpha>0$
be real constants. Let $P_{j}\in\Psi_{\delta}^{m+j\alpha}$, $j\in\mathbb{N}$,
be a sequence of symbols. We say that a symbol $P_{\hbar}\in\Psi_{\delta}^{m}$
admits the \textbf{asymptotic expansion} \[
P_{\hbar}\left(x,\xi\right)\sim\sum_{j=0}^{\infty}P_{j}\left(x,\xi,\hbar\right)\]
if for each integer $J$, one has \[
P_{\hbar}\left(x,\xi\right)-\sum_{j=0}^{J-1}P_{j}\left(x,\xi,\hbar\right)\in\Psi_{\delta}^{m+J\alpha}.\]
In case $\delta<1$ and $\alpha=1-\delta$, $P_{\hbar}$ is called
a \textbf{$\boldsymbol{\delta}$-classical} \textbf{symbol}. 
\end{defn}
We point out that in general these asymptotic expansions are not unique
since each term $P_{j}$ necessarily depends on $\hbar$. On the other
hand, one knows also that they are not convergent in general. Nevertheless,
one can apply the Borel resummation process for these symbols, as
stated in the following proposition (see e.g. \cite{roy_1}.)

\begin{lem}
\label{lem_borel_somme}Let $\delta\geq0$, $m$ and $\alpha>0$ be
real constants. For any sequence of symbols $P_{j}\in\Psi_{\delta}^{m+j\alpha}$,
there exists a symbol $P\in\Psi_{\delta}^{m}$ admitting the asymptotic
expansion $P\sim\sum P_{j}$.
\end{lem}
On the other hand, a slight modification of Borel construction yields
the following result.

\begin{lem}
\label{lem_borel_suite}Let $m$ and $\alpha>0$ be real constants.
For any (non-convergent) sequence of unitary operators $\hat{U}_{n}\in\mathcal{L}\left(L^{2}\right)$
satisfying \[
\left\Vert \hat{U}_{n}-\hat{U}_{n-1}\right\Vert _{\mathcal{L}\left(L^{2}\right)}=O\left(\hbar^{m+\alpha n}\right),\]
there exists an unitary operator $\hat{U}$ satisfying \[
\left\Vert \hat{U}-\hat{U}_{n-1}\right\Vert _{\mathcal{L}\left(L^{2}\right)}=O\left(\hbar^{m+\alpha n}\right).\]

\end{lem}

\subsubsection{Composition and commutators}

There is a composition law for the previously defined class of symbols
$\Psi_{\delta}^{m}$ provided $\delta<1$.

\begin{defn}
\label{def_moyal_product}Let $A_{\hbar},B_{\hbar}\in\Psi_{\delta}^{0}$.
We define their \textbf{(left) Moyal product} $A_{\hbar}\# B_{\hbar}$
by \[
A_{\hbar}\# B_{\hbar}\left(x,\xi\right)=\frac{1}{\left(2\pi\right)^{d}}\int_{\mathcal{T}}dy\sum_{k\in\Lambda^{*}}e^{ik\left(x-y\right)}A_{\hbar}\left(x,\xi+\hbar k\right)B_{\hbar}\left(y,\xi\right).\]

\end{defn}
\begin{lem}
\label{lem_moyal}Let $\delta\in\left[0,1\right)$. Let $\hat{A}$
and $\hat{B}$ be two PDOs in the class $\hat{\Psi}_{\delta}^{0}$
with symbols $A_{\hbar}$ and $B_{\hbar}$. Then, the product $\hat{C}=\hat{A}\hat{B}$
is a PDO in the same class and its symbol $C_{\hbar}$ is equal to
the Moyal product $C_{\hbar}=A_{\hbar}\# B_{\hbar}$. Moreover, the
symbol $C_{\hbar}$ admits the following $\delta$-classical asymptotic
expansion \[
A_{\hbar}\# B_{\hbar}\sim\sum_{j=0}^{\infty}C_{j}\left(\hbar\right),\]
 where the $C_{j}\in\Psi_{\delta}^{j\left(1-\delta\right)}$ are given
by \[
C_{j}\left(x,\xi,\hbar\right)=\left(\frac{\hbar}{i}\right)^{j}\sum_{\left|\alpha\right|=j}\frac{1}{\alpha!}\partial_{\xi}^{\alpha}A_{\hbar}\left(x,\xi\right)\partial_{x}^{\alpha}B_{\hbar}\left(x,\xi\right).\]

\end{lem}
From the previous lemma, one can easily obtain the symbol $A_{\hbar}\# B_{\hbar}-B_{\hbar}\# A_{\hbar}$
of the commutator $\left[\hat{A},\hat{B}\right]$ of two PDOs. In
case one of the two operators is in the class $\Psi_{0}^{0}$ (i.e.
with $\delta=0$) and does not depend on $x$, one has a slightly
better expansion that will be useful in the construction of the normal
form in the next section. 

\begin{lem}
\label{lem_commutateur_OPD}Let $\delta\in\left[0,1\right)$. Let
$A_{\hbar}\left(\xi\right)\in\Psi_{0}^{0}$ be a symbol independant
of $x$ and $B_{\hbar}\left(x,\xi\right)\in\Psi_{\delta}^{0}$ some
symbol. Then the commutator $C_{\hbar}=A_{\hbar}\# B_{\hbar}-B_{\hbar}\# A_{\hbar}$
is in the class $\Psi_{\delta}^{1}$ and admits an asymptotic expansion
of the following form \[
C_{\hbar}\left(x,\xi\right)\sim\frac{\hbar}{i}\left\{ A,B\right\} +\sum_{j=2}^{\infty}C_{j}\left(x,\xi,\hbar\right),\]
where the $C_{j}\in\Psi_{\delta}^{j}$ are given by \[
C_{j}\left(x,\xi,\hbar\right)=\left(\frac{\hbar}{i}\right)^{j}\sum_{\left|\gamma\right|=j}\frac{1}{\gamma!}\partial_{\xi}^{\gamma}A_{\hbar}\left(\xi\right)\partial_{x}^{\gamma}B_{\hbar}\left(x,\xi\right)\]
and where the asymptotic equivalence $\sim$ means that for each $J\in\mathbb{N}$,
one has \[
C_{\hbar}\left(x,\xi\right)-\sum_{j=0}^{J-1}C_{j}\left(x,\xi,\hbar\right)\in\Psi_{\delta}^{J}.\]

\end{lem}

\subsubsection{$L^{2}$ continuity and adjoints}

One can easily check that PDOs in the class $\hat{\Psi}_{\delta}^{m}$
are continuous from $C^{\infty}\left(\mathcal{T}\right)$ to $C^{\infty}\left(\mathcal{T}\right)$.
Moreover, the fact that the symbols together with all its derivatives
are uniformly bounded for $\left(x,\xi\right)\in T^{*}\mathcal{T}$
implies that the Calder\'on-Vaillancourt's theorem still holds in
the class $\hat{\Psi}_{\delta}^{0}$.

\begin{lem}
\label{lem_calderon_vaillancourt}Each PDO $\hat{P}\in\hat{\Psi}_{\delta}^{0}$
is continuous from $L^{2}\left(\mathcal{T}\right)$ to $L^{2}\left(\mathcal{T}\right)$.
Moreover, its norm is bounded by \[
\left\Vert \hat{P}\right\Vert _{\mathcal{L}\left(L^{2}\left(\mathcal{T}\right)\right)}\leq C\sup_{\left|\gamma\right|\leq\frac{d+1}{2}}\sup_{x,\xi}\left|\partial_{x}^{\gamma}P_{\hbar}\left(x,\xi\right)\right|.\]

\end{lem}
Let's turn now to the description of the symbol of an adjoint of a
PDO. 

\begin{lem}
\label{lem_adjoint_PDO}Let $\delta\in\left[0,1\right)$. For each
$\hat{P}\in\hat{\Psi}_{\delta}^{0}$, the adjoint $\hat{P}^{*}$ is
a PDO in the same class $\hat{\Psi}_{\delta}^{0}$ and its symbol,
denoted by $P_{\hbar}^{*}$, is given by \[
P_{\hbar}^{*}\left(x,\xi\right)=\sum_{k\in\Lambda^{*}}\frac{1}{\left(2\pi\right)^{d}}\int_{\mathcal{T}}dye^{ik\left(x-y\right)}\bar{P}_{\hbar}\left(y,\xi+\hbar k\right)\]
and admits the following $\delta$-classical asymptotic expansion
$\sum_{j=0}^{\infty}P_{j}^{*}\left(x,\xi,\hbar\right)$ where the
$P_{j}^{*}\in\Psi_{\delta}^{j\left(1-\delta\right)}$ are given by
\[
P_{j}^{*}\left(x,\xi,\hbar\right)=\left(\frac{\hbar}{i}\right)^{j}\sum_{\left|\gamma\right|=j}\frac{1}{\gamma!}\partial_{x}^{\gamma}\partial_{\xi}^{\gamma}\bar{P}_{\hbar}\left(x,\xi\right).\]

\end{lem}
For convenience, we say that $P_{\hbar}^{*}$ is the \textbf{adjoint
of the symbol} $P_{\hbar}$. Moreover, a straightforward calculation
shows that the Fourier series of the adjoint is given by the following
expression.

\begin{lem}
\label{lem_adjoint_fourier}Let $P_{\hbar}\in\Psi_{\delta}^{0}$ be
a symbol and $P_{\hbar}^{*}\in\Psi_{\delta}^{0}$ its adjoint. Then
their Fourier series are related as follows\[
\widetilde{P_{\hbar}^{*}}\left(k,\xi\right)=\overline{\tilde{P}_{\hbar}}\left(-k,\xi+\hbar k\right).\]

\end{lem}

\subsubsection{Exponentials and conjugacies}

We want to define the exponential $e^{i\hat{P}}$ for PDOs $\hat{P}$
in the class $\hat{\Psi}_{\delta}^{m}$, even for negative $m$. Lemma
\ref{lem_calderon_vaillancourt} insures that $\hat{P}$ is bounded
in $L^{2}\left(\mathcal{T}\right)$, but not uniformly in $\hbar$,
since its norm is of order $\hbar^{m}$. Nevertheless, one can define
the exponential thru the resolvent formula \[
e^{i\hat{P}}=\frac{1}{2\pi i}\int_{\mathcal{C_{\hbar}}}e^{iz}\left(z-\hat{P}\right)^{-1}dz,\]
provided $\mathcal{C}_{\hbar}\subset\mathbb{C}$ is a \emph{}cycle
surrounding the spectrum of $\hat{P}$, which is bounded for each
$\hbar$. When $m<0$, this exponential is not a PDO in the class
$\hat{\Psi}_{\delta}^{m}$. Nevertheless, for any $m$ it has the
usual properties, namely it is unitary in $L^{2}\left(\mathcal{T}\right)$
and satisfies $\frac{1}{i}\frac{d}{d\varepsilon}e^{i\varepsilon\hat{P}}=\hat{P}e^{i\varepsilon\hat{P}}=e^{i\varepsilon\hat{P}}\hat{P}$.

We remark in passing that for non-negative $m$, $e^{i\hat{P}}$ might
be in the class $\hat{\Psi}_{\delta}^{0}$ only up to an element in
$\hat{\Psi}_{\delta}^{\infty}$. The problem comes from the fact that
for PDO's on the torus, the resolvent $\left(z-\hat{P}\right)^{-1}$
itsef may be a PDO only up to an element in $\hat{\Psi}_{\delta}^{\infty}$.
We refer to \cite{roy_1} for a discussion of this issue.

Despite $e^{i\hat{P}}$ might not be in the class $\hat{\Psi}_{\delta}^{0}$,
the conjugacy $\hat{C}=e^{i\hat{P}}\hat{B}e^{-i\hat{P}}$ will be,
up to an negligible element, provided $m>-1$.

\begin{lem}
\label{lem_conjugaison_exponentielle}Let $m>-1$ and $0\leq\delta<\min\left(1,1+m\right)$.
Let $\hat{P}\in\hat{\Psi}_{\delta}^{m}$ and $\hat{B}\in\hat{\Psi}_{\delta}^{0}$
be two PDOs and let us consider the conjugacy $\hat{C}=e^{i\hat{P}}\hat{B}e^{-i\hat{P}}$.
Then $\hat{C}$ admits the following asymptotic expansion $\hat{C}\sim\sum_{n=0}^{\infty}\hat{C}_{n}$,
where $\hat{C}_{n}\in\Psi_{\delta}^{\left(m+1-\delta\right)n}$ is
given by \[
\hat{C}_{n}=\frac{i^{n}}{n!}\left[\right.\underbrace{\hat{P},...,\left[\hat{P}\right.}_{n},\left.\hat{B}\right]\left....\right].\]
The asymptotic expansion means that, for each integer $N\geq0$, the
remainder of the truncated series verifies $\hat{C}-\sum_{n=0}^{N-1}C_{n}\cong\hat{R}_{N}$
with $\hat{R}_{N}\in\hat{\Psi}_{\delta}^{\left(m+1-\delta\right)N}$.
\end{lem}
When the operator $\hat{B}$ is in $\hat{\Psi}_{0}^{0}$ (i.e. with
$\delta=0$) and its symbol does not depend on the $x$ variable,
then one can get a slightly better estimate (the gain is a factor
$\hbar^{\delta}$), that will be usefull subsequently.

\begin{lem}
\label{lem_conjugaison_exponentielle_2}Suppose now that $\hat{B}$
is in $\hat{\Psi}_{0}^{0}$ and that its symbol $B_{\hbar}\left(\xi\right)$
does not depend on $x$. Then the PDO's $\hat{C}_{n}$ of the asymptotic
expansion are in $\hat{C}_{n}\in\Psi_{\delta}^{\left(m+1-\delta\right)n+\delta}$,
and the remainders $\hat{R}_{N}\in\hat{\Psi}_{\delta}^{\left(m+1-\delta\right)N+\delta}$.
\end{lem}

\section{Quasi-resonant normal form}

\subsection{\label{sec_geom_resonance}Geometry of resonances}

\subsubsection{Nondegenerate Hamiltonians and resonances}

Denote by $\mathcal{B}=\mathbb{R}_{\xi}^{d}$ the momentum space.
A classical CI Hamiltonian $H\left(\xi\right)$ generates a linear
dynamics on each torus $\mathcal{T}_{\xi}$ that can be periodic,
ergodic or also partially (in a sub-torus) ergodic. For each $\xi\in\mathcal{B}_{\xi}$,
the resonant lattice of $dH$ at the point $\xi$ is defined by \[
\mathcal{R}_{\xi}=\left\{ k\in\Lambda^{*};dH_{\xi}\left(k\right)=0\right\} ,\]
where $\Lambda^{*}$ is the Fourier lattice. We thus have the following
cases :\\
$\bullet$ $\dim\mathcal{R}_{\xi}=0$ : We say that $\xi$ (or $\mathcal{T}_{\xi}$)
is \textbf{non-resonant}. The dynamics induced by $H$ is ergodic.\\
$\bullet$ $\dim\mathcal{R}_{\xi}>0$ : We say that $\xi$ (or $\mathcal{T}_{\xi}$)
is \textbf{resonant}. In this case, the dynamics is partially ergodic,
i.e. ergodic in a sub-torus of dimension $d-\dim\mathcal{R}_{\xi}$.
In particular, when $\dim\mathcal{R}_{\xi}=d-1$, we say that $\xi$
(or $\mathcal{T}_{\xi}$) is \textbf{periodic}.

From now on, the functions $dH_{\xi}\left(k\right)$ will be used
to define the resonances and their neighborhoods.

\begin{defn}
\label{def_ND_surface_sigma}For each non-vanishing $k\in\Lambda^{*}$,
we define the fonction $\Omega_{k}\in C^{\infty}\left(\mathcal{B}\right)$
by $\Omega_{k}\left(\xi\right)=dH_{\xi}\left(k\right)$ and the associated
\textbf{resonance surface} $\Sigma_{k}\subset\mathcal{B}$ by\[
\Sigma_{k}=\left\{ \xi\in\mathcal{B};\Omega_{k}\left(\xi\right)=0\right\} .\]

\end{defn}
The resonant set $\Sigma_{k}$ will indeed be a hyper-surface as soon
as we will impose $H$ to be \emph{non-degenerate}. Such a condition
is a very common assumption in K.A.M. like or Nekhoroshev like theories
which insures that the CI dynamics {}``varies enough'' from one
torus to another one. The nondegeneracy condition that we will use
is slightly weaker than Kolmogorov's one \cite{kolmogorov} or Arnol'd's
one \cite{arnold_3} and equivalent to Bryuno's one \cite{bruno}.
See \cite{roy_5} for a review of the nondegeneracy conditions.

\begin{defn}
\label{def_ND_condition}A CI Hamiltonian $H\left(\xi\right)$ is
said to be \textbf{non-degenerate} if for each non-vanishing $k\in\Lambda^{*}$
and each point $\xi\in\Sigma_{k}$, we have $d\left(\Omega_{k}\right)_{\xi}\neq0$. 
\end{defn}
This implies that the set $\Sigma_{k}$ is a codimension 1 submanifold
of $\mathcal{B}$ and thus deserves its name {}``resonance surface''.
Moreover, if $k_{1},...,k_{n}\in\Lambda^{*}$ are linearly independent,
then one can show (see e.g. \cite{roy_1}) that the submanifolds $\Sigma_{k_{j}}$
are transverse.

\subsubsection{Quasi-resonant blocks\label{sec_quasi_reson_block}}

The first step in the construction of the announced quasi-resonant
normal form, is to obtain a covering of the momentum space $\mathcal{B}$
by regions attached to each particular kind of resonance. For each
resonant torus $\mathcal{T}_{\xi}$, we consider a {}``small'' neighborhood
and we remove from it a {}``sufficiently large'' neighborhood of
higher order resonances, as Pöschel did in \cite{poeschel_2}, in
order to get the so-called \emph{quasi-resonant blocks.} One the other
hand, in our semi-classical context, one needs to let both notions
{}``small'' and {}``sufficiently large'' depend on $\hbar$. We
now elaborate on Pöschel's construction, yet incorporating $\hbar$
in the right place. For this, we will fix two exponents $\gamma>0$
and $\delta>0$ which control respectively the {}``amount'' of resonances
we consider and the {}``size'' of the quasi-resonant zones. 

\begin{defn}
\label{def_resonance_lattice}A n-dimensional sub-lattice $\mathcal{R}$
of the Fourier lattice $\Lambda^{*}$ is called a \textbf{resonance}
$\boldsymbol{\hbar^{-\gamma}}$\textbf{-lattice} (or simply a \textbf{}$\boldsymbol{\hbar^{-\gamma}}$\textbf{-lattice})
if there exists a basis $\left(e_{1},...,e_{n}\right)$ of $\mathcal{R}$
such that $\left|e_{j}\right|\leq\hbar^{-\gamma}$ for all $j=1..n$.
\end{defn}
Similarly with Definition \ref{def_ND_surface_sigma}, we define the
resonant manifold attached to each $\hbar^{-\gamma}$-lattice $\mathcal{R}$,
making use of the function $\Omega_{k}$ previously defined.

\begin{defn}
\label{def_resonance_manifold}For each resonance $\hbar^{-\gamma}$-lattice
$\mathcal{R}$, the associated \textbf{resonance manifold} $\Sigma_{\mathcal{R}}\subset\mathcal{B}$
is defined by\[
\Sigma_{\mathcal{R}}=\left\{ \xi\in\mathcal{B};\forall k\in\mathcal{R}\Rightarrow\Omega_{k}\left(\xi\right)=0\right\} .\]
For consistency of the notations, in the case of the trivial lattice
$\mathcal{R}=0$, we define $\Sigma_{0}$ to be the whole $\mathcal{B}$.
\end{defn}
For a given $\hbar^{-\gamma}$-lattice $\mathcal{R}$, the resonant
manifold $\Sigma_{\mathcal{R}}$ is thus the set of points $\xi$
at which the resonant lattice of $dH$ is exactely equal to $\mathcal{R}$.
Moreover, we obviously have $\Sigma_{\mathcal{R}}=\bigcap_{k\in\mathcal{R}}\Sigma_{k}$
and the notation is still consistent when $\mathcal{R}$ is 1-dimensional,
i.e. of the form $\mathcal{R}=\mathbb{Z}.k_{0}$, if we write $\Sigma_{\mathbb{Z}.k_{0}}=\Sigma_{k_{0}}$.
As mentionned above, the nondegeneracy hypothesis implies that for
each $\hbar^{-\gamma}$-lattice $\mathcal{R}$ of dimension $n$,
the manifold $\Sigma_{\mathcal{R}}$ is of codimension $n$ in $\mathcal{B}$.

\begin{defn}
\label{def_resonance_zone}For each resonance $\hbar^{-\gamma}$-lattice
$\mathcal{R}$ of dimension $n>0$, the associated \textbf{resonance
zone} $\mathcal{Z}_{\mathcal{R}}\subset\mathcal{B}$ is defined by\[
\mathcal{Z}_{\mathcal{R}}=\left\{ \xi\in\mathcal{B};\forall X\in\mathbb{R}.\mathcal{R}\Rightarrow\frac{\left|\Omega_{X}\left(\xi\right)\right|}{\left|X\right|}<\frac{2^{n}\hbar^{\delta-\gamma n}}{\textrm{vol}\left(\mathcal{R}\right)}\right\} .\]
We also define $\mathcal{Z}_{0}=\mathcal{B}$.
\end{defn}
The denominator $\textrm{vol}\left(\mathcal{R}\right)$ in the previous
definition refers to the volume of a fundamental domain of the lattice
$\mathcal{R}$. 

\begin{defn}
\label{def_resonance_zone*}We denote by $\mathcal{Z}_{n}^{*}$ the
union of all resonance zones of order $n$. For $0\leq n\leq d$,
we have \[
\mathcal{Z}_{n}^{*}=\bigcup_{\dim\mathcal{R}=n}\mathcal{Z}_{\mathcal{R}}.\]
We call $\mathcal{Z}_{n}^{*}$ the \textbf{zone of} $\boldsymbol{n}$-\textbf{resonances}. \textbf{}
\end{defn}
Then we remove, from each resonance zone, a neighborhood of all next
order resonances and obtain the so-called resonance blocks.

\begin{defn}
\label{def_resonance_block}For each resonance $\hbar^{-\gamma}$-lattice
$\mathcal{R}$ of dimension $n$, the associated \textbf{resonance
block} $\mathcal{B}_{\mathcal{R}}\subset\mathcal{B}$ is defined by\[
\mathcal{B}_{\mathcal{R}}=\mathcal{Z}_{\mathcal{R}}\setminus\mathcal{Z}_{n+1}^{*},\]
where we defined $\mathcal{Z}_{d+1}^{*}=\emptyset$ for consistency
of the notations. We also denote \[
\mathcal{B}_{n}^{*}=\bigcup_{\dim\mathcal{R}=n}\mathcal{B}_{\mathcal{R}}\]
the \textbf{block of} $\boldsymbol{n}$-\textbf{resonances} and call
$\mathcal{B}_{0}^{*}=\mathcal{B}_{0}$ the \textbf{non-resonant block}. 
\end{defn}
\begin{center}\includegraphics[%
  scale=0.88]{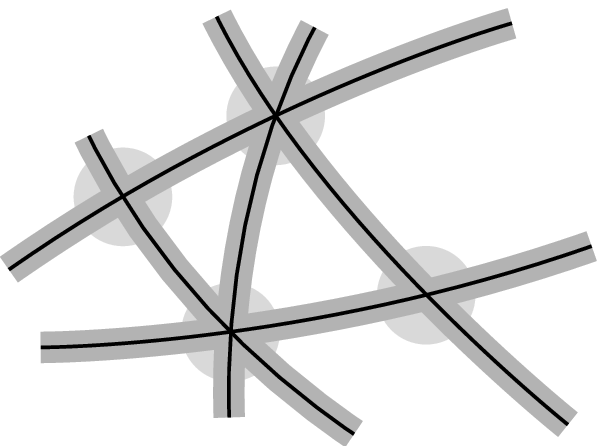}\end{center}

The picture above represents the situation in dimension $d=2$. The
black lines are the resonance manifolds for $1$-dimensional resonance
lattices (i.e. the set of periodic tori). They intersect on resonance
manifolds of $2$-dim{\-}ensional resonance lattices. The dark grey
regions represent the associated zones of $1$-resonances and the
light grey regions are the zones of $2$-resonances. This picture
can also be understood as a $2$-dimensional cross section of $\mathcal{B}$
in dimension $d=3$. 

The resonant zones are defined in such a way (with sizes increasing
with the order) that all points $\xi$ in a given block $\mathcal{B}_{\mathcal{R}}$
are {}``almost resonant'' for all $k\in\mathcal{R}$ and not {}``almost
resonant'' for all $k\notin\mathcal{R}$. The precise statement of
this assertion is given in the following lemma. We refer the reader
to Pöschel's article (\cite{poeschel_2}, p. 201) for the proof.

\begin{lem}
\label{lem_geometric_lemma}For each resonance $\hbar^{-\gamma}$-lattice
$\mathcal{R}$ of dimension $0\leq n<d$ and each $\xi\in\mathcal{B_{\mathcal{R}}}$,
we have \[
\forall k\notin\mathcal{R},\left|k\right|\leq\hbar^{-\gamma}\Rightarrow\frac{\left|\Omega_{k}\left(\xi\right)\right|}{\left|k\right|}\geq\hbar^{\delta}.\]
 This formula still holds for the non-resonant block $\mathcal{B}_{0}$.
\end{lem}
On the other hand, the resonance blocks form a covering of the space
$\mathcal{B}$ since they satisfy \[
\mathcal{B}=\mathcal{B}_{0}^{*}\cup\mathcal{B}_{1}^{*}...\cup\mathcal{B}_{d}^{*}.\]

\subsection{\label{sec_quasi_reson_averaging}Quasi-resonant averaging}

\subsubsection{Resonant averaging}

For any $n$-dimensional resonance lattice $\mathcal{R}$, one can
consider the averaging of functions along $\mathcal{R}^{°}$, the
dual space of $\mathcal{R}$, in the following way.

\begin{defn}
\label{def_average_resonant}For any function $f\in C^{\infty}\left(T^{*}\mathcal{T},\mathbb{C}\right)$
and any $n$-dimensional resonance lattice $\mathcal{R}$, we call
the \textbf{average of $f$ with respect to $\mathcal{R}$}, or the
$\mathcal{R}$\textbf{-average of $f$,} the function $\mathcal{R}\textrm{-av}\left(f\right)\in C^{\infty}\left(T^{*}\mathcal{T},\mathbb{C}\right)$
defined by\[
\mathcal{R}\textrm{-av}\left(f\right)\left(x,\xi\right)=\int_{0}^{1}dt_{1}...\int_{0}^{1}dt_{d-n}\, f\left(x+t_{1}X_{1}+...+t_{d-n}X_{d-n},\xi\right),\]
 where $\left(X_{1},...,X_{d-n}\right)$ is any basis of $\mathcal{R}^{°}\subset\Lambda$.
In particular, we will note $\left\langle \left\langle f\right\rangle \right\rangle =\left\{ 0\right\} \textrm{-av}\left(f\right)\left(\xi\right)$
the average of $f$ over the whole torus.
\end{defn}
One can easily check that this definition does not depend on the choice
of the basis $\left(X_{1},...,X_{d-n}\right)$. Moreover, it is easy
to show that the Fourier series of such an averaged function has the
simple form given below.

\begin{lem}
\label{lem_average_fourier}Let $\mathcal{R}$ be any resonance lattice.
Let $f\in C^{\infty}\left(T^{*}\mathcal{T},\mathbb{C}\right)$ be
any function and $\left\langle f\right\rangle =\mathcal{R}\textrm{-av}\left(f\right)$
its $\mathcal{R}$-average. If we denote by $\widetilde{f}$ the Fourier
series of $f$ then $\widetilde{\left\langle f\right\rangle }$, the
Fourier series of $\left\langle f\right\rangle $, is given by the
following expression. \[
\widetilde{\left\langle f\right\rangle }\left(k,\xi\right)=\left\{ \begin{array}{c}
\widetilde{f}\left(k,\xi\right)\textrm{ for }k\in\mathcal{R}\textrm{ }\\
0\textrm{ for }k\notin\mathcal{R}.\end{array}\right.\]
In particular, the Fourier series of $\left\langle \left\langle f\right\rangle \right\rangle $
verifies $\widetilde{\left\langle \left\langle f\right\rangle \right\rangle }\left(0,\xi\right)=\widetilde{f}\left(0,\xi\right)$
and vanishes for $k\neq0$.
\end{lem}

\subsubsection{Quasi-resonant averaging}

Let us consider the previously defined covering of $\mathcal{B}$
with resonant blocks $\mathcal{B}_{\mathcal{R}}$. For any symbol
$K_{\hbar}\in\Psi_{\delta}^{m}$, we will construct a symbol in $\Psi_{\delta}^{m}$
which is an $\mathcal{R}$-averaged function in each blocks $\mathcal{B}_{\mathcal{R}}$
and which is moreover exactly the $\mathcal{R}$-average of $K_{\hbar}$
on $\Sigma_{\mathcal{R}}\cap\mathcal{B_{R}}$. For the construction,
we need a truncature function that will be keeped fixed and that will
localize in a neighborhood of size $\hbar^{\delta}$ of the resonant
blocks. Precisely, let us choose a function $\chi\in C_{0}^{\infty}\left(\mathbb{R}\right)$
with value in $\left[0,1\right]$, symmetric, vanishing for $\left|t\right|\geq1$
and such that $\chi-1$ is flat at $t=0$.

\begin{defn}
\label{def_quasi_average}For any $\delta>0$ and any symbol $K_{\hbar}\in\Psi_{\delta}^{m}$,
we define $A_{\hbar}$ the \textbf{$\boldsymbol{\hbar^{\delta}}$-average}
of $K_{\hbar}$ by \[
\tilde{A}_{\hbar}\left(k,\xi\right)=\chi\left(\frac{\Omega_{k}\left(\xi\right)}{\left|k\right|\hbar^{\delta}}\right)\tilde{K}_{\hbar}\left(k,\xi\right)\]
for all $\xi\in\mathcal{B}$ and all non-vanishing $k\in\Lambda^{*}$,
and $\tilde{A}_{\hbar}\left(0,\xi\right)=\tilde{K}_{\hbar}\left(0,\xi\right)$
for all $\xi\in\mathcal{B}$.
\end{defn}
The following lemma tells us that the \textbf{$\hbar^{\delta}$}-average
has the property that for each resonance $\hbar^{-\gamma}$-lattice
$\mathcal{R}$, on the resonant manifold $\Sigma_{\mathcal{R}}\cap\mathcal{B_{R}}$
the \textbf{$\hbar^{\delta}$}-average of $K_{\hbar}\in\Psi_{\delta}^{m}$
is equal (up to $O\left(\hbar^{\infty}\right)$) to the $\mathcal{R}$-average
of $K_{\hbar}$, and in the resonant block $\mathcal{B_{R}}$ it is
a $\mathcal{R}$-averaged function.

\begin{prop}
\label{prop_quasi_average}The \textbf{$\hbar^{\delta}$}-average
$A_{\hbar}$ of any symbol $K_{\hbar}\in\Psi_{\delta}^{m}$ is in
the class $\Psi_{\delta}^{m}$ and has the following properties. For
each resonance $\hbar^{-\gamma}$-lattice $\mathcal{R}$, we have
:\[
A_{\hbar}-\mathcal{R}\textrm{-av}\left(K_{\hbar}\right)\in\Psi_{\delta}^{\infty}\left(\mathcal{T}\times\left(\Sigma_{\mathcal{R}}\cap\mathcal{B_{R}}\right)\right)\,\,\,\textrm{ and \,}\,\, A_{\hbar}-\mathcal{R}\textrm{-av}\left(A_{\hbar}\right)\in\Psi_{\delta}^{\infty}\left(\mathcal{T}\times\mathcal{B_{R}}\right).\]

\end{prop}
\begin{proof}
Let us first show that $A_{\hbar}$ is indeed in the class $\Psi_{\delta}^{m}$.
First of all, for each multi-index $\beta\in\mathbb{N}^{d}$, the
derivative of the Fourier series of $A_{\hbar}$ is given by \[
\partial_{\xi}^{\beta}\tilde{A}_{\hbar}\left(k,\xi\right)=\sum_{\beta^{'}\leq\beta}C_{\beta^{'}}^{\beta}\left(\partial_{\xi}^{\beta-\beta^{'}}\widetilde{K}_{\hbar}\left(k,\xi\right)\right)\partial_{\xi}^{\beta^{'}}\left(\chi\left(\frac{\Omega_{k}\left(\xi\right)}{\left|k\right|\hbar^{\delta}}\right)\right)\]
for non-vanishing $k\in\Lambda^{*}$ and simply $\partial_{\xi}^{\beta}\tilde{A}_{\hbar}\left(0,\xi\right)=\partial_{\xi}^{\beta}\tilde{K}_{\hbar}\left(0,\xi\right)$
for $k=0$. On the other hand, according to Lemma \ref{lem_OPD_fourier_symbole},
the fact that $K_{\hbar}$ is in $\Psi_{\delta}^{m}$ implies for
all $s$ the estimates\[
\left|\partial_{\xi}^{\beta-\beta^{'}}\tilde{K}_{\hbar}\left(k,\xi\right)\right|\leq C\left(s,\beta-\beta^{'}\right)\frac{\hbar^{m-\delta\left|\beta-\beta^{'}\right|}}{\left(1+\left|k\right|^{2}\right)^{\frac{s}{2}}},\]
where $C\left(s,\beta-\beta^{'}\right)$ is a positive constant. However,
one gets easily convinced that the derivatives of the function $\chi$
are of the form \[
\left|\partial_{\xi}^{\beta^{'}}\left(\chi\left(\frac{\Omega_{k}\left(\xi\right)}{\left|k\right|\hbar^{\delta}}\right)\right)\right|\leq\sum_{n=1}^{\left|\beta^{'}\right|}c\left(n\right)\hbar^{-\delta n}\chi^{\left(n\right)}\left(\frac{\Omega_{k}\left(\xi\right)}{\left|k\right|\hbar^{\delta}}\right),\]
where the constants $c\left(n\right)$ depend only on $H$ and its
derivatives. We then notice that $\hbar^{-\delta n}\leq\hbar^{-\delta\left|\beta^{'}\right|}$
and that all the derivatives $\chi^{\left(n\right)}$ are bounded
(thanks to the fact that $\chi^{\left(n\right)}\in C_{0}^{\infty}\left(\mathbb{R}\right)$),
what implies the estimate \[
\left|\partial_{\xi}^{\beta^{'}}\left(\chi\left(\frac{\Omega_{k}\left(\xi\right)}{\left|k\right|\hbar^{\delta}}\right)\right)\right|\leq\hbar^{-\delta\left|\beta^{'}\right|}C\left(\beta^{'}\right)\]
 for all $k$, all $\hbar$ and all $\xi$. This shows that \[
\left|\partial_{\xi}^{\beta}\tilde{A}_{\hbar}\left(k,\xi\right)\right|\leq C^{'}\left(s,\beta\right)\frac{\hbar^{m-\delta\left|\beta-\beta^{'}\right|}\hbar^{-\delta\left|\beta^{'}\right|}}{\left(1+\left|k\right|^{2}\right)^{\frac{s}{2}}}=C^{'}\left(s,\beta\right)\frac{\hbar^{m-\delta\left|\beta\right|}}{\left(1+\left|k\right|^{2}\right)^{\frac{s}{2}}},\]
where $C^{'}\left(s,\beta\right)$ is a positive constant. Using again
Lemma \ref{lem_OPD_fourier_symbole}, we deduce that $A_{\hbar}$
is a symbol in the class $\Psi_{\delta}^{m}$. 

Let us now prove that $A_{\hbar}$ is an $\mathcal{R}$-averaged function,
up to $O\left(\hbar^{\infty}\right)$, in each resonant block $\mathcal{B}_{\mathcal{R}}$.
For this, let's define the remainder $R_{\hbar}=A_{\hbar}-\mathcal{R}\textrm{-av}\left(A_{\hbar}\right)$,
which is in the class $\Psi_{\delta}^{m}$ since $A_{\hbar}$ and
$\mathcal{R}\textrm{-av}\left(A_{\hbar}\right)$ are. The Fourier
series of $R_{\hbar}$ is thus given by $\widetilde{R}_{\hbar}\left(k,\xi\right)=\chi\left(\frac{\Omega_{k}\left(\xi\right)}{\left|k\right|\hbar^{\delta}}\right)\tilde{K}_{\hbar}\left(k,\xi\right)$
for $k\in\mathcal{R}$ and $\widetilde{R}_{\hbar}\left(k,\xi\right)=0$
for $k\in\mathcal{R}$. For each $\beta\in\mathbb{N}^{d}$ we now
estimate $\partial_{\xi}^{\beta}\tilde{R}_{\hbar}\left(k,\xi\right)$
at each point $\xi\in\mathcal{B_{R}}$. For all $k\in\mathcal{R}$
with $\left|k\right|\leq\hbar^{-\gamma}$, one simply has $\partial_{\xi}^{\beta}\tilde{R}_{\hbar}\left(k,\xi\right)=0$
everywhere. For all $k\notin\mathcal{R}$ with $\left|k\right|\leq\hbar^{-\gamma}$,
one has $\left|\frac{\Omega_{k}\left(\xi\right)}{\left|k\right|\hbar^{\delta}}\right|\geq1$
at each point $\xi\in\mathcal{B_{R}}$ thanks to Lemma \ref{lem_geometric_lemma}.
The truncature function $\chi\left(t\right)$ vanishing for $t\geq1$,
it follows that $\partial_{\xi}^{\beta}\tilde{R}_{\hbar}\left(k,\xi\right)=0$
for all $k\notin\mathcal{R}$ with $\left|k\right|\leq\hbar^{-\gamma}$
and all $\xi\in\mathcal{B_{R}}$. Finally, for all $k$ with $\left|k\right|>\hbar^{-\gamma}$,
we simply use the fact that $R_{\hbar}\in\Psi_{\delta}^{m}\left(\mathcal{T}\right)$,
what implies (Lemma \ref{lem_OPD_fourier_symbole}) that \begin{equation}
\left|\partial_{\xi}^{\beta}\tilde{R}_{\hbar}\left(k,\xi\right)\right|\leq C\left(s,\beta\right)\frac{\hbar^{m-\delta\left|\beta\right|}}{\left(1+\left|k\right|^{2}\right)^{\frac{s}{2}}},\label{eq_moyenne_quasi_inter_1}\end{equation}
where $C\left(s,\beta\right)$ is a positive constant.

Reconstructing the Fourier series of $R_{\hbar}$, one then gets for
each $\alpha\in\mathbb{N}^{d}$\[
\partial_{x}^{\alpha}\partial_{\xi}^{\beta}R_{\hbar}\left(x,\xi\right)=\sum_{\left|k\right|>\hbar^{-\gamma}}e^{ikx}i^{\left|\alpha\right|}k^{\alpha}\partial_{\xi}^{\beta}\tilde{R}_{\hbar}\left(k,\xi\right),\]
and thus for each $\xi\in\mathcal{B_{R}}$ one has\begin{eqnarray*}
\left|\partial_{x}^{\alpha}\partial_{\xi}^{\beta}R_{\hbar}\left(x,\xi\right)\right| & \leq & C\left(s,\beta\right)\hbar^{m-\delta\left|\beta\right|}\sum_{\left|k\right|>\hbar^{-\gamma}}\frac{\left|k\right|^{\left|\alpha\right|}}{\left(1+\left|k\right|^{2}\right)^{\frac{s}{2}}}\\
 & \leq & C\left(\alpha,\beta,N\right)\hbar^{m-\delta\left|\beta\right|}\hbar^{\gamma N},\end{eqnarray*}
where we have defined $N$ by $s=\left|\alpha\right|+d+N$. This estimate
holds for each $s$ and thus for each $N$. This implies that $R_{\hbar}\in\Psi_{\delta}^{\infty}\left(\mathcal{T}\times\mathcal{B_{R}}\right)$.

Let's now turn to the second property, namely on $\Sigma_{\mathcal{R}}\cap\mathcal{B_{R}}$,
$A_{\hbar}$ is equal up to $O\left(\hbar^{\infty}\right)$ to $\mathcal{R}\textrm{-av}\left(K_{\hbar}\right)$.
To prove this, we define another remainder $S_{\hbar}=A_{\hbar}-\mathcal{R}\textrm{-av}\left(K_{\hbar}\right)$
whose Fourier series is given by\[
\widetilde{S}_{\hbar}\left(k,\xi\right)=\left\{ \begin{array}{c}
\left(\chi\left(\frac{\Omega_{k}\left(\xi\right)}{\left|k\right|\hbar^{\delta}}\right)-1\right)\tilde{K}_{\hbar}\left(k,\xi\right)\textrm{ for }k\in\mathcal{R}\\
\chi\left(\frac{\Omega_{k}\left(\xi\right)}{\left|k\right|\hbar^{\delta}}\right)\tilde{K}_{\hbar}\left(k,\xi\right)\textrm{ for }k\notin\mathcal{R}\end{array}\right.\]
 As before, we will estimate $\partial_{\xi}^{\beta}\tilde{S}_{\hbar}\left(k,\xi\right)$
for each $\beta\in\mathbb{N}^{d}$ at each point $\xi\in\mathcal{B_{R}}$.
For all $k\in\mathcal{R}$ with $\left|k\right|\leq\hbar^{-\gamma}$,
one has $\Omega_{k}\left(\xi\right)=0$ at each point $\xi\in\Sigma_{\mathcal{R}}\cap\mathcal{B_{R}}$
by definition of $\Sigma_{k}$. On the other hand, the function $\chi\left(t\right)-1$
is flat at $t=0$. Thus, for all $k\in\mathcal{R}$ one has $\partial_{\xi}^{\beta}\tilde{S}_{\hbar}\left(k,\xi\right)=0$
at each $\xi\in\Sigma_{\mathcal{R}}\cap\mathcal{B_{R}}$. For the
cases $k\notin\mathcal{R}$ with $\left|k\right|\leq\hbar^{-\gamma}$
and $k\notin\mathcal{R}$ with $\left|k\right|>\hbar^{-\gamma}$,
we argue as before and obtain that $S_{\hbar}\in\Psi_{\delta}^{\infty}\left(\mathcal{T}\times\left(\Sigma_{\mathcal{R}}\cap\mathcal{B_{R}}\right)\right)$.
\end{proof}
The \textbf{$\hbar^{\delta}$}-average $A_{\hbar}$ of a symbol $K_{\hbar}$
has nevertheless a drawback : it is not self-adjoint even when $K_{\hbar}$
is. To solve this, we have to show that the self-adjoint part $\frac{1}{2}\left(A_{\hbar}+A_{\hbar}^{*}\right)$
has the same average properties as $A_{\hbar}$ has. 

\begin{prop}
\label{prop_quasi_average_self_adjoint}Let the parameters $\gamma$
and $\delta$ satisfy $\gamma+\delta<1$. Let $K_{\hbar}\in\Psi_{\delta}^{m}$
be any self-adjoint symbol and $A_{\hbar}\in\Psi_{\delta}^{m}$ its
\textbf{$\hbar^{\delta}$}-average as in Definition \ref{def_quasi_average}.
Then the adjoint $A_{\hbar}^{*}\in\Psi_{\delta}^{m}$ has the same
properties as $A_{\hbar}$ (as given in Proposition \ref{prop_quasi_average}),
as well as its self-adjoint part $\frac{1}{2}\left(A_{\hbar}+A_{\hbar}^{*}\right)$.
\end{prop}
\begin{proof}
According to Lemma \ref{lem_adjoint_fourier}, the Fourier series
of $A_{\hbar}^{*}$ is given by \begin{eqnarray*}
\tilde{A}_{\hbar}\left(k,\xi\right) & = & \overline{\tilde{A}_{\hbar}}\left(-k,\xi+\hbar k\right)\\
 & = & \overline{\chi}\left(\frac{\Omega_{-k}\left(\xi+\hbar k\right)}{\left|-k\right|\hbar^{\delta}}\right)\overline{\tilde{K}_{\hbar}}\left(-k,\xi+\hbar k\right),\end{eqnarray*}
for all nonvanishing $k$ and simply $\tilde{A}_{\hbar}^{*}\left(0,\xi\right)=\overline{\tilde{K}_{\hbar}}\left(0,\xi\right)$.
Using then the facts that $K_{\hbar}$ is self-adjoint, that $\Omega_{-k}=-\Omega_{k}$
and that the function $\chi$ is real and symetric, we obtain \[
\tilde{A}_{\hbar}^{*}\left(k,\xi\right)=\chi\left(\frac{\Omega_{k}\left(\xi+\hbar k\right)}{\left|k\right|\hbar^{\delta}}\right)\tilde{K}_{\hbar}\left(k,\xi\right),\]
for all non-vanishing $k$ and $\tilde{A}_{\hbar}^{*}\left(0,\xi\right)=\tilde{K}_{\hbar}\left(0,\xi\right)$.
We now prove that $A_{\hbar}^{*}-\mathcal{R}\textrm{-av}\left(A_{\hbar}^{*}\right)\in\Psi_{\delta}^{\infty}\left(\mathcal{T}\times\mathcal{B_{R}}\right)$.
We introduce the remainder $R_{\hbar}=A_{\hbar}^{*}-\mathcal{R}\textrm{-av}\left(A_{\hbar}^{*}\right)$,
which is proved to be in the class $\Psi_{\delta}^{m}$ in the same
way as we proved it for $A_{\hbar}$ itself. The Fourier series of
$R_{\hbar}$ is given by $\widetilde{R}_{\hbar}\left(k,\xi\right)=\chi\left(\frac{\Omega_{k}\left(\xi+\hbar k\right)}{\left|k\right|\hbar^{\delta}}\right)\tilde{K}_{\hbar}\left(k,\xi\right)$
for all $k\notin\mathcal{R}$ and $0$ otherwise. 

For each $\beta\in\mathbb{N}^{d}$ we have to estimate $\partial_{\xi}^{\beta}\tilde{R}_{\hbar}\left(k,\xi\right)$
at each point $\xi\in\mathcal{B_{R}}$. For all $k\in\mathcal{R}$
with $\left|k\right|\leq\hbar^{-\gamma}$, one simply has $\partial_{\xi}^{\beta}\tilde{R}_{\hbar}\left(k,\xi\right)=0$
everywhere. For all $k\notin\mathcal{R}$ with $\left|k\right|\leq\hbar^{-\gamma}$,
one has $\left|\frac{\Omega_{k}\left(\xi\right)}{\left|k\right|\hbar^{\delta}}\right|\geq1$
at each point $\xi\in\mathcal{B_{R}}$ thanks to Lemma \ref{lem_geometric_lemma}.
Nevertheless, we have to evaluate $\Omega_{k}$ at $\xi+\hbar k$
and not at $\xi$. However, the bound $\left|k\right|\leq\hbar^{-\gamma}$
insures that $\frac{\Omega_{k}\left(\xi+\hbar k\right)}{\left|k\right|\hbar^{\delta}}=\frac{\Omega_{k}\left(\xi\right)}{\left|k\right|\hbar^{\delta}}+O\left(\hbar^{1-\delta-\gamma}\right)$.
The relation between $\delta$ and $\gamma$ then implies that $\hbar^{1-\delta-\gamma}$
is small when $\hbar\rightarrow0$, and using the fact that the truncature
function $\chi\left(t\right)$ is flat when $\left|t\right|\geq1$,
we obtain $\chi\left(\frac{\Omega_{k}\left(\xi+\hbar k\right)}{\left|k\right|\hbar^{\delta}}\right)=O\left(\hbar^{\left(1-\delta-\gamma\right)\infty}\right)=O\left(\hbar^{\infty}\right)$.
A similar argument for the derivatives $\chi^{\left(n\right)}$ yields
$\partial_{\xi}^{\beta}\tilde{R}_{\hbar}\left(k,\xi\right)=O\left(\hbar^{-\left|\beta\right|\delta}\hbar^{\left(1-\delta-\gamma\right)\infty}\right)=O\left(\hbar^{\infty}\right)$
for all $k\notin\mathcal{R}$ with $\left|k\right|\leq\hbar^{-\gamma}$
and all $\xi\in\mathcal{B_{R}}$. Finally, for all $k$ with $\left|k\right|>\hbar^{-\gamma}$,
we use as before the fact that $R_{\hbar}\in\Psi_{\delta}^{m}\left(\mathcal{T}\right)$,
what implies (Lemma \ref{lem_OPD_fourier_symbole}) that \[
\left|\partial_{\xi}^{\beta}\tilde{R}_{\hbar}\left(k,\xi\right)\right|\leq C\left(s,\beta\right)\frac{\hbar^{m-\delta\left|\beta\right|}}{\left(1+\left|k\right|^{2}\right)^{\frac{s}{2}}},\]
where $C\left(s,\beta\right)$ is a positive constant.

Reconstructing the Fourier series of $R_{\hbar}$, one gets for each
$\alpha\in\mathbb{N}^{d}$\[
\partial_{x}^{\alpha}\partial_{\xi}^{\beta}R_{\hbar}\left(x,\xi\right)=\sum_{\left|k\right|\leq\hbar^{-\gamma}}e^{ikx}i^{\left|\alpha\right|}k^{\alpha}\partial_{\xi}^{\beta}\tilde{R}_{\hbar}\left(k,\xi\right)+\sum_{\left|k\right|>\hbar^{-\gamma}}e^{ikx}i^{\left|\alpha\right|}k^{\alpha}\partial_{\xi}^{\beta}\tilde{R}_{\hbar}\left(k,\xi\right),\]
and thus for each $\xi\in\mathcal{B_{R}}$ one has\[
\left|\partial_{x}^{\alpha}\partial_{\xi}^{\beta}R_{\hbar}\left(x,\xi\right)\right|\leq O\left(\hbar^{\infty}\right)\sum_{\left|k\right|\leq\hbar^{-\gamma}}\left|k\right|^{\left|\alpha\right|}+C\left(s,\beta\right)\hbar^{m-\delta\left|\beta\right|}\sum_{\left|k\right|>\hbar^{-\gamma}}\frac{\left|k\right|^{\left|\alpha\right|}}{\left(1+\left|k\right|^{2}\right)^{\frac{s}{2}}}.\]
This holds for all $s$ and thus one has $\left|\partial_{x}^{\alpha}\partial_{\xi}^{\beta}R_{\hbar}\left(x,\xi\right)\right|=O\left(\hbar^{\infty}\right)$.
This proves that $R_{\hbar}\in\Psi_{\delta}^{\infty}\left(\mathcal{T}\times\mathcal{B_{R}}\right)$.

We now let the reader check that, following the same arguments, one
can show that $A_{\hbar}^{*}-\mathcal{R}\textrm{-av}\left(K_{\hbar}^{*}\right)\in\Psi_{\delta}^{\infty}\left(\mathcal{T}\times\Sigma_{\mathcal{R}}\right)$.
This shows that $A_{\hbar}^{*}$ has the same average properties as
$A_{\hbar}$, as well as $\frac{1}{2}\left(A_{\hbar}+A_{\hbar}^{*}\right)$.
\end{proof}
From now on, the function $A_{\hbar}$ in Definition \ref{def_quasi_average}
is called the \textbf{$\boldsymbol{\hbar^{\delta}}$-average} of $K_{\hbar}$
and the one in Proposition \ref{prop_quasi_average_self_adjoint}
is called the \textbf{self-adjoint $\boldsymbol{\hbar^{\delta}}$-average}
of $K_{\hbar}$.

\subsection{\label{sec_NF}Semi-classical normal form}

Let's now turn to the study of the \emph{homological equation} arising
at each step of the construction of the normal form given in Theorem
\ref{theo_quasi_resonant_NF}.

\begin{prop}
[Homological equation]\label{prop_homological_eq}Let $H\left(\xi\right)\in\Psi^{0}$
be a non-degenerate CI Hamiltonian. Let $K_{\hbar}\in\Psi_{\delta}^{m}$
be any symbol and $A_{\hbar}\in\Psi_{\delta}^{m}$ its \textbf{$\hbar^{\delta}$}-average.
Then there exists a symbol $P_{\hbar}\in\Psi_{\delta}^{m-\delta}$
solution of the equation\[
\left\{ P_{\hbar},H\right\} +K_{\hbar}-A_{\hbar}=0.\]

\end{prop}
\begin{proof}
We first write the Fourier series of the homological equation, i.e.\begin{equation}
i\Omega_{k}\left(\xi\right)\tilde{P}_{\hbar}\left(k,\xi\right)+\tilde{K}_{\hbar}\left(k,\xi\right)-\tilde{A}_{\hbar}\left(k,\xi\right)=0.\label{eq_lem_hom_eq_inter_1}\end{equation}
For $k=0$, the equation is fullfilled since $\Omega_{0}\left(\xi\right)=0$
and $\tilde{A}_{\hbar}\left(0,\xi\right)=\tilde{K}_{\hbar}\left(0,\xi\right)$.
We can choose for example $\tilde{P}_{\hbar}\left(k,\xi\right)=0$.
For all $k\neq0$, the Fourier series of the \textbf{$\hbar^{\delta}$}-average
is given by $\tilde{A}_{\hbar}\left(k,\xi\right)=\chi\left(\frac{\Omega_{k}\left(\xi\right)}{\left|k\right|\hbar^{\delta}}\right)\tilde{K}_{\hbar}\left(k,\xi\right)$.
We then notice that the function $\phi\left(t\right)=\frac{1-\chi\left(t\right)}{t}$
is smooth. This implies that the function $P_{\hbar}\left(x,\xi\right)$
defined by \[
\tilde{P}_{\hbar}\left(k,\xi\right)=\frac{i\tilde{K}_{\hbar}\left(k,\xi\right)}{\left|k\right|\hbar^{\delta}}\phi\left(\frac{\Omega_{k}\left(\xi\right)}{\left|k\right|\hbar^{\delta}}\right)\]
is well-defined and satisfies Equation \ref{eq_lem_hom_eq_inter_1}.
Moreover, proceeding as in Proposition \ref{prop_quasi_average} for
proving that the $\hbar^{\delta}$-average is a symbol in the class
$\Psi_{\delta}^{m}$, and using the smoothness of $\phi$, one shows
that for all $\alpha,\beta\in\mathbb{N}^{d}$, 

\[
\left|\partial_{x}^{\alpha}\partial_{\xi}^{\beta}P_{\hbar}\left(x,\xi\right)\right|\leq C\left(\alpha,\beta\right)\hbar^{m-\delta-\delta\left|\beta\right|}.\]
This proves that $P_{\hbar}\in\Psi_{\delta}^{m-\delta}$. 
\end{proof}
\begin{thm}
[Semi-classical normal form]\label{theo_quasi_resonant_NF}Let us
consider a PDO $\hat{H}\in\hat{\Psi}^{0}$ with non-degenerate CI
symbol $H\left(\xi\right)\in\Psi^{0}$ and any self-adjoint perturbation
$\hbar^{\kappa}\hat{K}_{0}\in\hat{\Psi}_{\delta}^{\kappa}$, with
$\kappa>0$. Let us choose small parameters $\gamma\geq0$ and $\delta\geq0$
such that $\delta<1-\gamma$ and $\delta<\frac{\kappa}{3}$, and let
us consider the covering of $\mathcal{B}$ with resonance blocks as
described in Section \ref{sec_quasi_reson_block}. 

Then there exist an unitary operator $\hat{U}$ and a self-adjoint
PDO $\hat{K}\in\hat{\Psi}_{\delta}^{0}$ satisfying \begin{equation}
\hat{U}\left(\hat{H}+\hbar^{\kappa}\hat{K}_{0}\right)\hat{U}^{-1}=\hat{H}+\hbar^{\kappa}\hat{A}+O\left(\hbar^{\infty}\right),\label{eq_FN_quasi_resonant}\end{equation}
where the symbol of $\hat{A}\in\hat{\Psi}_{\delta}^{0}$ is the self-adjoint
\textbf{$\hbar^{\delta}$}-average of the symbol $K_{\hbar}$. Moreover,
one has $\hat{U}=\mathbb{I}+O\left(\hbar^{\kappa-1-\delta}\right)$
and $\hat{K}=\hat{K}_{0}+O\left(\hbar^{\alpha}\right)$, with $\alpha=\min\left(1-\delta,\kappa-3\delta\right)$.
\end{thm}
\begin{proof}
The exponent $\alpha$ is positive because of the restrictions on
$\delta$ and $\kappa$. We first prove that there exist self-adjoint
PDOs $\hat{P}_{0}\in\hat{\Psi}_{\delta}^{-\delta}\left(\mathcal{T}\right)$
and $\hat{K}_{1}\in\hat{\Psi}_{\delta}^{\alpha}\left(\mathcal{T}\right)$
such that \begin{equation}
e^{i\hbar^{\kappa-1}\hat{P}_{0}}\left(\hat{H}+\hbar^{\kappa}\hat{K}_{0}\right)e^{-i\hbar^{\kappa-1}\hat{P}_{0}}\cong\hat{H}+\hbar^{\kappa}\hat{A}_{0}+\hbar^{\kappa}\hat{K}_{1},\label{eq_FN_quasi_resonant_inter_1}\end{equation}
where $A_{0}\left(\hbar\right)\in\Psi_{\delta}^{0}$ is the \textbf{}self-adjoint
\textbf{$\hbar^{\delta}$}-average of $K_{0}\left(\hbar\right)$.

\noindent $\bullet$ Indeed, Lemma \ref{lem_conjugaison_exponentielle_2}
tells us that \begin{eqnarray*}
e^{i\hbar^{\kappa-1}\hat{P}_{0}}\hat{H}e^{-i\hbar^{\kappa-1}\hat{P}_{0}} & \cong & \hat{H}+i\hbar^{\kappa-1}\left[\hat{P}_{0},\hat{H}\right]+O\left(\hbar^{2\left(\kappa-1-\delta+1-\delta\right)+\delta}\right),\\
 & \cong & \hat{H}+i\hbar^{\kappa-1}\left[\hat{P}_{0},\hat{H}\right]+O\left(\hbar^{2\kappa-3\delta}\right).\end{eqnarray*}
 On the other hand, one can apply Lemma \ref{lem_commutateur_OPD}
which insures that the symbol of $\left[\hat{P}_{0},\hat{H}\right]$
is equal to $\frac{\hbar}{i}\left\{ P_{0},H\right\} +O\left(\hbar^{2-\delta}\right)$.
This yields \begin{eqnarray*}
e^{i\hbar^{\kappa-1}\hat{P}_{0}}\hat{H}e^{-i\hbar^{\kappa-1}\hat{P}_{0}} & \cong & \hat{H}+i\hbar^{\kappa-1}\left(\frac{\hbar}{i}\widehat{\left\{ P_{0},H\right\} }\right)+O\left(\hbar^{1+\kappa-\delta}\right)+O\left(\hbar^{2\kappa-3\delta}\right)\\
 & \cong & \hat{H}+\hbar^{\kappa}\widehat{\left\{ P_{0},H\right\} }+O\left(\hbar^{\kappa+\alpha}\right),\end{eqnarray*}
where we have used the previously defined parameter $\alpha$.

\noindent $\bullet$ Similarly, Lemma \ref{lem_conjugaison_exponentielle}
tells us that \begin{eqnarray*}
e^{i\hbar^{\kappa-1}\hat{P}_{0}}\hbar^{\kappa}\hat{K}_{0}e^{-i\hbar^{\kappa-1}\hat{P}_{0}} & \cong & \hbar^{\kappa}\hat{K}_{0}+O\left(\hbar^{\kappa}\hbar^{\kappa-1-\delta+1-\delta}\right).\\
 & \cong & \hbar^{\kappa}\hat{K}_{0}+O\left(\hbar^{\kappa+\alpha}\right),\end{eqnarray*}
where we have used that $\kappa-2\delta>\kappa-3\delta\geq\alpha$
and thus $\hbar^{\kappa-2\delta}\ll\hbar^{\alpha}$. 

We then write the symbol of Equation (\ref{eq_FN_quasi_resonant_inter_1})
and, dividing by $\hbar^{\kappa}$, one sees that we have to solve
\begin{equation}
\left\{ P_{0},H\right\} +K_{0}-A_{0}=O\left(\hbar^{\alpha}\right).\label{eq_FN_quasi_resonant_inter_2}\end{equation}
 Actually, Lemma \ref{prop_homological_eq} insures that we can find
a symbol $P_{0}^{'}\in$ $\Psi_{\delta}^{-\delta}$ such that we have
exactely $\left\{ P_{0}^{'},H\right\} +K_{0}-A_{0}^{'}=0$, with $A_{0}^{'}\left(\hbar\right)\in\Psi_{\delta}^{0}$
the \textbf{$\hbar^{\delta}$}-average of $K_{0}\left(\hbar\right)$.
Nevertheless, neither $A_{0}^{'}$ nor $P_{0}^{'}$ are self-adjoint.
The adjoint of this equation is $\left\{ P_{0}^{'},H\right\} ^{*}+K_{0}-\left(A_{0}^{'}\right)^{*}=0$
since $K_{0}$ is self-adjoint. Moreover, using Lemma \ref{lem_moyal}
and Lemma \ref{lem_commutateur_OPD}, one sees that $\left\{ P_{0}^{'},H\right\} ^{*}=\left\{ \left(P_{0}^{'}\right)^{*},H\right\} +O\left(\hbar^{1-\delta}\right)$.
This implies that $P_{0}=\frac{1}{2}\left(P_{0}^{'}+\left(P_{0}^{'}\right)^{*}\right)$
satisfies Equation (\ref{eq_FN_quasi_resonant_inter_2}) with $A_{0}=\frac{1}{2}\left(A_{0}^{'}+\left(A_{0}^{'}\right)^{*}\right)$
being the self-adjoint \textbf{$\hbar^{\delta}$}-average of $K_{0}\left(\hbar\right)$,
thanks to Proposition \ref{prop_quasi_average_self_adjoint} and to
the fact that $\hbar^{1-\delta}\leq\hbar^{\alpha}$. The quantized
of $P_{0}$ thus satisfies Equation (\ref{eq_FN_quasi_resonant_inter_1})
with a self-adjoint $\hat{K}_{1}\in\hat{\Psi}_{\delta}^{\alpha}$.
If we define $\hat{V}_{0}=e^{i\hbar^{\kappa-1}\hat{P}_{0}}$, we have
\begin{equation}
\hat{V}_{0}\left(\hat{H}+\hbar^{\kappa}\hat{K}_{0}\right)\hat{V}_{0}^{-1}\cong\hat{H}+\hbar^{\kappa}\hat{A}_{0}+\hbar^{\kappa}\hat{K}_{1}.\label{eq_FN_quasi_resonant_etape_1}\end{equation}

This equation is the initial step of the following induction process.
Let us suppose that at the step $n\geq0$, we have found self-adjoint
PDOs $\hat{K}_{1},...,\hat{K}_{n+1}$, with $\hat{K}_{j}\in\hat{\Psi}_{\delta}^{j\alpha}$
and unitary operators $\hat{V}_{0},...,\hat{V}_{n}$ such that \begin{equation}
\hat{V}_{n}...\hat{V}_{0}\left(\hat{H}+\hbar^{1}\hat{K}\right)\hat{V}_{0}^{-1}...\hat{V}_{n}^{-1}\cong\hat{H}+\hbar^{\kappa}\sum_{j=0}^{n}\hat{A}_{j}+\hbar^{\kappa}\hat{K}_{n+1},\label{eq_FN_quasi_resonant_n}\end{equation}
where $A_{j}\left(\hbar\right)\in\Psi_{\delta}^{j\alpha}$ is the
self-adjoint \textbf{$\hbar^{\delta}$}-average of $K_{j}\left(\hbar\right)$.
Then we look for a PDO $\hat{K}_{n+2}\in\hat{\Psi}_{\delta}^{\left(n+2\right)\alpha}$
and a unitary operator$\hat{V}_{n+1}$ satisfying \begin{equation}
\hat{V}_{n+1}\hat{V}_{n}...\hat{V}_{0}\left(\hat{H}+\hbar^{\kappa}\hat{K}\right)\hat{V}_{0}^{-1}...\hat{V}_{n}^{-1}\hat{V}_{n+1}^{-1}\cong\hat{H}+\hbar^{\kappa}\sum_{j=0}^{n+1}\hat{A}_{j}+\hbar^{\kappa}\hat{K}_{n+2},\label{eq_FN_quasi_resonant_n1}\end{equation}
where $A_{n+1}\left(\hbar\right)\in\Psi_{\delta}^{\left(n+1\right)\alpha}$
is the self-adjoint \textbf{$\hbar^{\delta}$}-average of $K_{n+1}\left(\hbar\right)$.
Looking for $\hat{V}_{n+1}$ of the form $\hat{V}_{n+1}=e^{i\hbar^{\kappa-1}\hat{P}_{n+1}}$,
with $\hat{P}_{n+1}\in\hat{\Psi}_{\delta}^{-\delta+\left(n+1\right)\alpha}$
self-adjoint, and inserting Equation (\ref{eq_FN_quasi_resonant_n})
into Equation (\ref{eq_FN_quasi_resonant_n1}), we obtain\begin{equation}
e^{i\hbar^{\kappa-1}\hat{P}_{n+1}}\left[\hat{H}+\hbar^{\kappa}\sum_{j=0}^{n}\hat{A}_{j}+\hbar^{\kappa}\hat{K}_{n+1}\right]e^{-i\hbar^{\kappa-1}\hat{P}_{n+1}}\cong\hat{H}+\hbar^{\kappa}\sum_{j=0}^{n+1}\hat{A}_{j}+\hbar^{\kappa}\hat{K}_{n+2}.\label{eq_FN_quasi_resonant_etape_n2}\end{equation}
We now apply Lemmas \ref{lem_conjugaison_exponentielle} and \ref{lem_conjugaison_exponentielle_2}
for each term inside the bracket $\left[\,\right]$. 

\noindent $\bullet$ First, Lemma \ref{lem_conjugaison_exponentielle_2}
tells us that \begin{eqnarray*}
\hat{V}_{n+1}\left[\hat{H}\right]\hat{V}_{n+1}^{-1} & \cong & \hat{H}+i\hbar^{\kappa-1}\left[\hat{P}_{n+1},\hat{H}\right]+O\left(\hbar^{2\left(\kappa-1-\delta+\left(n+1\right)\alpha+1-\delta\right)+\delta}\right)\\
 & \cong & \hat{H}+i\hbar^{\kappa-1}\left[\hat{P}_{n+1},\hat{H}\right]+O\left(\hbar^{\kappa+\left(n+2\right)\alpha}\right),\end{eqnarray*}
were we have used $\hbar^{\kappa-3\delta}\leq\hbar^{\alpha}$ and
$\hbar^{\left(2n+3\right)\alpha}\ll\hbar^{\left(n+2\right)\alpha}$
provided $n\geq0$.

\noindent $\bullet$ On the other hand, one can apply Lemma \ref{lem_commutateur_OPD}
which insures that the symbol of $\left[\hat{P}_{n+1},\hat{H}\right]$
equals to $\frac{\hbar}{i}\left\{ P_{n+1},H\right\} +O\left(\hbar^{2-\delta+\left(n+1\right)\alpha}\right)$,
i.e.\[
\frac{\hbar}{i}\left\{ P_{n+1},H\right\} +O\left(\hbar^{1+\left(n+2\right)\alpha}\right),\]
where we have used $\hbar^{1-\delta}\leq\hbar^{\alpha}$. Therefore,
the symbol of $\hat{V}_{n+1}\left[\hat{H}\right]\hat{V}_{n+1}^{-1}$
is\[
H+\hbar^{\kappa}\left\{ P_{n+1},H\right\} +O\left(\hbar^{\kappa+\left(n+2\right)\alpha}\right).\]

\noindent $\bullet$ Then, Lemma \ref{lem_conjugaison_exponentielle}
provides, for all $j=0..n$,\begin{eqnarray*}
\hat{V}_{n+1}\left[\hbar^{\kappa}\hat{A}_{j}\right]\hat{V}_{n+1}^{-1} & \cong & \hbar^{\kappa}\hat{A}_{j}+O\left(\hbar^{\kappa}\hbar^{j\alpha}\hbar^{\kappa-1-\delta+\left(n+1\right)\alpha+1-\delta}\right)\\
 & \cong & \hbar^{\kappa}\hat{A}_{j}+O\left(\hbar^{\kappa+\left(n+2\right)\alpha}\right).\end{eqnarray*}
 where we have used $\hbar^{\kappa-2\delta}\ll\hbar^{\kappa-3\delta}\leq\hbar^{\alpha}$
and $\hbar^{j\alpha}\leq1$.

\noindent $\bullet$ Finally, Lemma \ref{lem_conjugaison_exponentielle}
yields \begin{eqnarray*}
\hat{V}_{n+1}\left[\hbar^{\kappa}\hat{K}_{n+1}\right]\hat{V}_{n+1}^{-1} & \cong & \hbar^{\kappa}\hat{K}_{n+1}+O\left(\hbar^{\kappa}\hbar^{\left(n+1\right)\alpha}\hbar^{\kappa-1-\delta+\left(n+1\right)\alpha+1-\delta}\right)\\
 & \cong & \hbar^{\kappa}\hat{K}_{n+1}+O\left(\hbar^{\kappa+\left(2n+3\right)\alpha}\right)\\
 & \cong & \hbar^{\kappa}\hat{K}_{n+1}+O\left(\hbar^{\kappa+\left(n+2\right)\alpha}\right),\end{eqnarray*}
 where we have used $\hbar^{\kappa-2\delta}\ll\hbar^{\alpha}$ and
$\hbar^{\left(2n+3\right)\alpha}\ll\hbar^{\left(n+2\right)\alpha}$
provided $n\geq0$.

If we consider these different estimates, if we take the symbol of
Equation (\ref{eq_FN_quasi_resonant_etape_n2}) and if we divide by
$\hbar^{\kappa}$, we see that we have to solve\begin{equation}
\left\{ P_{n+1},H\right\} +K_{n+1}-A_{n+1}=O\left(\hbar^{\left(n+2\right)\alpha}\right),\label{eq_FN_quasi_resonant_inter_3}\end{equation}
 where $A_{n+1}\left(\hbar\right)$ is the self-adjoint \textbf{$\hbar^{\delta}$}-average
of $K_{n+1}\left(\hbar\right)$. Then we use Proposition \ref{prop_homological_eq}
which insures that we can find a symbol $P_{n+1}^{'}\in\Psi_{\delta}^{-\delta+\left(n+1\right)\alpha}$,
such that we have exactely $\left\{ P_{n+1}^{'},H\right\} +K_{n+1}-A_{n+1}^{'}=0$,
where $A_{n+1}^{'}\left(\hbar\right)$ is the \textbf{$\hbar^{\delta}$}-average
of $K_{n+1}\left(\hbar\right)$. Using the same technique as for the
initial step $n=0$, we show that $P_{n+1}=\frac{1}{2}\left(P_{n+1}^{'}+\left(P_{n+1}^{'}\right)^{*}\right)$
satisfies Equation (\ref{eq_FN_quasi_resonant_inter_3}) with $A_{n+1}=\frac{1}{2}\left(A_{n+1}^{'}+\left(A_{n+1}^{'}\right)^{*}\right)$
being the self-adjoint \textbf{$\hbar^{\delta}$}-average of $K_{n+1}\left(\hbar\right)$.
The quantized of $P_{n+1}$ thus satisfies Equation (\ref{eq_FN_quasi_resonant_etape_n2})
with a self-adjoint $\hat{K}_{n+2}\in\hat{\Psi}_{\delta}^{\left(n+2\right)\alpha}$.
This concludes the iterative process. 

Finally, we apply Borel's construction (Lemma \ref{lem_borel_suite})
to the sequence of unitary operators $\hat{U}_{n}=\hat{V}_{n}...\hat{V}_{0}$
and obtain an unitary operator $\hat{U}$ which satisfies $\left\Vert \hat{U}-\hat{U}_{n-1}\right\Vert _{\mathcal{L}\left(L^{2}\right)}=O\left(\hbar^{\kappa-1+\delta+\alpha n}\right)$
for each $n$. Similarly, Lemma \ref{lem_borel_somme} insures that
there exists a self-adjoint PDO $\hat{K}\in\hat{\Psi}_{\delta}^{0}$
verifying $\hat{K}\sim\sum_{n}\hat{K}_{n}$. Moreover, if we define
$A_{\hbar}$ as the self-adjoint \textbf{$\hbar^{\delta}$}-average
of the symbol $K_{\hbar}$, we can see that $\hat{A}\sim\sum_{n}\hat{A}_{n}$.
Therefore, we have proved that the operator $\hat{R}$ defined by
\[
\hat{U}\left(\hat{H}+\hbar^{\kappa}\hat{K}_{0}\right)\hat{U}^{-1}=\hat{H}+\hbar^{\kappa}\hat{A}+\hat{R},\]
 satisfies  $\left\Vert \hat{R}\right\Vert _{\mathcal{L}\left(L^{2}\right)}=O\left(\hbar^{\infty}\right)$.
\end{proof}

\section{\label{sec_quasimodes}Application : quasimodes}

As an example of application of the normal form given in Theorem \ref{theo_quasi_resonant_NF},
we can easily construct quasimodes associated with the block $\mathcal{B}_{0}$.

\begin{thm}
[Non-resonant quasimodes]\label{theo_quasimode_non_resonant}Let
$\hat{H}\in\hat{\Psi}^{0}$ be a PDO with non-degenerate CI symbol
$H\left(\xi\right)\in\Psi^{0}$ and $\hbar^{\kappa}\hat{K}_{0}\in\hat{\Psi}_{\delta}^{\kappa}$
any self-adjoint perturbation with $\kappa>0$. Let us fix small parameters
$\gamma\geq0$ and $\delta\geq0$ such that $\delta<1-\gamma$ and
$\delta<\frac{\kappa}{3}$, and consider the covering of $\mathcal{B}$
with resonance blocks.

Then for each familly $k_{\hbar}\in\Lambda^{*}$ such that $\hbar k_{\hbar}$
remains in the non-resonant block $\mathcal{B}_{0}$, there is a $\hbar^{\infty}$-quasimode
$\varphi_{\hbar}$ of the perturbed operator \[
\left\Vert \left(\hat{H}+\hbar^{\kappa}\hat{K}_{0}-E_{\hbar}\right)\varphi_{\hbar}\right\Vert =O\left(\hbar^{\infty}\right)\]
with the property $\varphi_{\hbar}=e^{ik_{\hbar}x}+O\left(\hbar^{\kappa-1-\delta}\right)$
and with quasi-eigenvalue\[
E_{\hbar}=H\left(\hbar k_{\hbar}\right)+\hbar^{\kappa}F_{\hbar}\left(\hbar k_{\hbar}\right),\]
where the $x$-independant symbol $F_{\hbar}\in\Psi_{\delta}^{0}$
is given by \[
F_{\hbar}\left(\hbar k_{\hbar}\right)=\left\langle \left\langle K_{0}\right\rangle \right\rangle \left(\hbar k_{\hbar}\right)+O\left(\hbar^{\alpha}\right),\]
with $\left\langle \left\langle K_{0}\right\rangle \right\rangle \left(\xi\right)$
being the average of $K_{0}$ over the torus and $\alpha$ being defined
by $\alpha=\min\left(1-\delta,\kappa-3\delta\right)$.
\end{thm}
\begin{center}\emph{\includegraphics[%
  width=5.5cm]{dessins/stable_ev.eps}}\end{center}

The picture above illustrates the case $d=2$. The grey region is
the zone $\mathcal{Z}_{1}^{*}$ of $1$-resonances and the dots stand
for the lattice $\hbar\mathbb{Z}^{d}$. 

\begin{proof}
Indeed, according to Theorem \ref{theo_quasi_resonant_NF}, the perturbed
operator is conjugate to its normal form $\widehat{NF}\cong\hat{H}+\hbar^{\kappa}\hat{A}$,
where $A_{\hbar}$ is the self-adjoint \textbf{$\hbar^{\delta}$}-average
of $K_{\hbar}$, which is a symbol satisfying $\hat{K}=\hat{K}_{0}+O\left(\hbar^{\alpha}\right)$.
Moreover, in the block $\mathcal{B}_{0}$, Proposition \ref{prop_quasi_average_self_adjoint}
tells us that $A_{\hbar}$ is simply $A_{\hbar}=\left\langle \left\langle K_{\hbar}\right\rangle \right\rangle +O\left(\hbar^{\infty}\right)$.
On the other hand, as the averaged symbol $\left\langle \left\langle K_{\hbar}\right\rangle \right\rangle $
is independent on $x$, the eigenvalues of $\hat{H}+\hbar^{\kappa}\widehat{\left\langle \left\langle K_{\hbar}\right\rangle \right\rangle }$
are given by $E_{k}\left(\hbar\right)=H\left(\hbar k\right)+\hbar^{\kappa}\left\langle \left\langle K_{\hbar}\right\rangle \right\rangle \left(\hbar k\right)$,
for each $k\in\Lambda^{*}$, and the associated eigenvectors are simply
the exponential functions $e^{ikx}$. These functions are thus also
$\hbar^{\infty}$-quasimodes of the $\widehat{NF}$ for each familly
$k_{\hbar}$ such that $\hbar k_{\hbar}\in\mathcal{B}_{0}$ for all
$\hbar$, and the quasi-eigenvalues are $E_{k_{\hbar}}\left(\hbar\right)$.
Then, applying the operators $\hat{U}$ which conjugate the perturbed
operator to $\widehat{NF}$, we obtain $\hbar^{\infty}$-quasimodes
$\varphi_{\hbar}=\hat{U}^{*}\left(e^{ik_{\hbar}x}\right)$ of the
perturbed operator, with the same quasi-eigenvalues. Finally, we notice
that according to the properties of $\hat{U}$, the quasimodes have
the form $\varphi_{\hbar}=e^{ik_{\hbar}x}+O\left(\hbar^{\kappa-1-\delta}\right)$.
Moreover, according to the expression of $K_{\hbar}$, the eigenvalues
have the expression\[
E_{k_{\hbar}}\left(\hbar\right)=H\left(\hbar k_{\hbar}\right)+\hbar^{\kappa}\left\langle \left\langle K_{0}\right\rangle \right\rangle \left(\hbar k_{\hbar}\right)+O\left(\hbar^{\kappa+\alpha}\right).\]

\end{proof}
These quasi-eigenvalues are very easily constructed but the number
we can construct depends on the size of the block $\mathcal{B}_{0}$
as $\hbar$ goes to zero. This size depends on the parameters $\gamma$
and $\delta$, which control respectively the amount of resonance
zones we consider and their width (see below).

\noindent \begin{center}\begin{minipage}[t]{0.50\textwidth}%
\noindent \begin{center}\includegraphics[%
  scale=0.9]{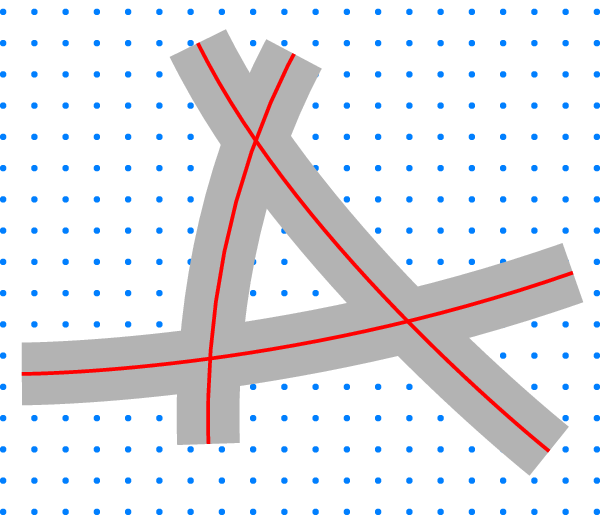}\\
''small'' $\delta$ and $\gamma$\end{center}\end{minipage}%
\begin{minipage}[t]{0.50\textwidth}%
\noindent \begin{center}\includegraphics[%
  scale=0.9]{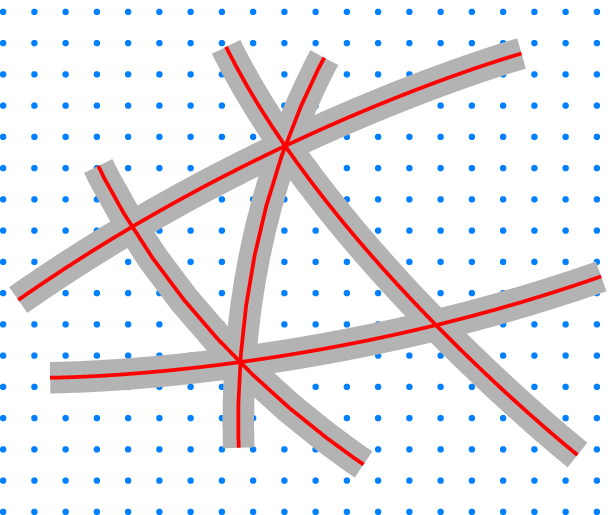}\\
''big'' $\delta$ and $\gamma$\end{center}\end{minipage}%
\end{center}

With an appropriate choice of $\gamma$ and $\delta$, one can insure
that the relative volume of $\mathcal{B}_{0}$ tends to $1$ as $\hbar$
goes to zero, as follows.

\begin{prop}
\label{prop_volume_bloc}Let $\gamma>0$ and $\delta>0$ such that
$\delta>2d\gamma$. For each ball $\mathcal{O}\subset\mathcal{B}$
and each $n=1..d$, the volume of the block of $n$-resonances $\mathcal{B}_{n}^{*}$
is of order \[
vol\left(\mathcal{B}_{n}^{*}\cap\mathcal{O}\right)=\left(\hbar^{n\left(\delta-2d\gamma\right)}\right).\]

\end{prop}
\begin{proof}
First, one can show (see e.g. \cite{roy_6} or \cite{roy_1}) that
the nondegeneracy condition for $H$ implies that there exist two
constants $T$ and $C$ such that for all $k\neq0$ and all $\xi$
satisfying $\frac{\Omega_{k}\left(\xi\right)}{\left|k\right|}\leq T$,
one has $\left|d\frac{\Omega_{k}\left(\xi\right)}{\left|k\right|}\right|\geq C$.
For $\hbar$ small enough ($\hbar^{\delta-\gamma n}\leq T$) and for
each $k\neq0$, the points $\xi$ in the resonant zone $\mathcal{Z}_{k}\cap\mathcal{O}$
satisfy $\left|d\frac{\Omega_{k}\left(\xi\right)}{\left|k\right|}\right|\geq C$.
The function $\frac{\Omega_{k}\left(\xi\right)}{\left|k\right|}$
is thus suitable to measure lengths and this implies that the volume
of the resonant zone $Z_{\mathcal{R}}\cap\mathcal{O}$, for any $n$-dimensional
lattice $\mathcal{R}$, is bounded by $vol\left(Z_{\mathcal{R}}\cap\mathcal{O}\right)=O\left(\left(\hbar^{\delta-\gamma n}\right)^{n}\right)$
uniformly with respect to $\mathcal{R}$. On the other hand, the volume
of $\mathcal{B}_{k}\cap\mathcal{O}$ is bounded in the same way since
$\mathcal{B}_{k}\subset\mathcal{Z}_{k}$. 

The bloc $\mathcal{B}_{1}^{*}\cap\mathcal{O}$ of $1$-resonances
is the union of the $\mathcal{B}_{k}\cap\mathcal{O}$ for all non-vanishing
$k$ with $\left|k\right|\leq\hbar^{-\gamma}$. If we bound the sum
over all primitive $k$ with $\left|k\right|\leq\hbar^{-\gamma}$
by the sum over all $k$ with $\left|k\right|\leq\hbar^{-\gamma}$,
we get the following estimate\[
vol\left(\mathcal{B}_{1}^{*}\cap\mathcal{O}\right)\leq C\hbar^{\delta-\gamma}\sum_{\left|k\right|\leq\hbar^{-\gamma}}=O\left(\hbar^{\delta-\gamma-d\gamma}\right).\]
Similarly, for $n=2..d$, the block $\mathcal{B}_{n}^{*}\cap\mathcal{O}$
of $n$-resonances is the union of the blocks $\mathcal{B}_{\mathcal{R}}\cap\mathcal{O}$
for all the resonance $\hbar^{-\gamma}$-lattices $\mathcal{R}$,
i.e. the sub-lattices which admit a basis $\left(e_{1},...,e_{n}\right)$
composed of vector with norm $\left|e_{j}\right|\leq\hbar^{-\gamma}$.
The volume of $\mathcal{B}_{n}^{*}\cap\mathcal{O}$ is then roughly
bounded by\[
vol\left(\mathcal{B}_{n}^{*}\cap\mathcal{O}\right)\leq C^{'}\hbar^{n\left(\delta-n\gamma\right)}\sum_{\begin{array}{c}
{\scriptstyle \left|e_{1}\right|\leq\hbar^{-\gamma}}\\
\vdots\\
{\scriptstyle \left|e_{n}\right|\leq\hbar^{-\gamma}}\end{array}}=O\left(\hbar^{n\left(\delta-n\gamma\right)-\gamma dn}\right).\]
This estimate is actually too rough when $n\geq\frac{d}{2}$ since,
for example, the number of lattices of $d$-resonances is equal to
$1$ rather than to $\hbar^{-\gamma d^{2}}$. Therefore, for $n\geq\frac{d}{2}$,
we will rather count the orthogonal lattices. This gives a number
of lattices of order $\hbar^{-\gamma d\left(d-n\right)}$ and thus
a volume of order $vol\left(\mathcal{B}_{n}^{*}\cap\mathcal{O}\right)=O\left(\hbar^{n\left(\delta-n\gamma\right)-\gamma d\left(d-n\right)}\right)$
for $n\geq\frac{d}{2}$. On the other hand, the inequality involving
$\gamma$ and $\delta$ implies the following :

\noindent $\bullet$ When $0<n<\frac{d}{2}$, we have \[
n\left(\delta-n\gamma\right)-\gamma dn>n\left(\delta-\frac{d}{2}\gamma-\gamma d\right)>n\left(\delta-2\gamma d\right).\]

\noindent $\bullet$ When $\frac{d}{2}\leq n\leq d$, we have $n\left(\delta-n\gamma\right)-\gamma d\left(d-n\right)=n\left(\delta-n\gamma+d\gamma-\gamma\frac{d^{2}}{n}\right)$
and the bracket is estimated by \[
\delta-n\gamma+d\gamma-\gamma\frac{d^{2}}{n}\geq\delta-d\gamma+d\gamma-\gamma\frac{2d^{2}}{d}=\delta-2\gamma d.\]

This thus shows that the volume $vol\left(\mathcal{B}_{n}^{*}\cap\mathcal{O}\right)$
is of order $O\left(\hbar^{n\left(\delta-2\gamma d\right)}\right)\ll1$
for all $n=1..d$.
\end{proof}
\bibliographystyle{amsplain}
\bibliography{/home/roy/math/biblio/biblio_nico}

\end{document}